\newif\ifarxiv
\arxivtrue 

\ifarxiv
\documentclass[11pt]{article}
\else
\documentclass[ba]{imsart}
\fi

\RequirePackage{amsthm,amsmath,amsfonts,amssymb}
\RequirePackage[authoryear]{natbib}
\RequirePackage[colorlinks,citecolor=blue,urlcolor=blue,backref=page,backref=page]{hyperref}
\RequirePackage{graphicx}

\ifarxiv
\else
\pubyear{2026}
\arxiv{TBA}
\volume{TBA}
\issue{TBA}
\firstpage{1}
\lastpage{1}

\startlocaldefs
\fi
\theoremstyle{plain}

\newtheorem{theorem}{Theorem}[section]
\newtheorem{lemma}[theorem]{Lemma}
\theoremstyle{definition}

\theoremstyle{remark}

\usepackage{multirow}
\newtheorem{prop}{Proposition}
\usepackage{subfig}
\usepackage[ruled,vlined]{algorithm2e}
\usepackage{tikz}
\usepackage{pgfplots}
\usepgfplotslibrary{fillbetween}
\pgfplotsset{compat=1.18}
\renewcommand{\Pr}{\mbox{Pr}}
\newcommand*{\Bagel}{Bagel~}
\usepackage{pdfpages}
\usetikzlibrary{arrows.meta,external}
\newcommand*\ab{.4}
\tikzset{
  net node/.style = {circle, minimum width=2*\ab cm, inner sep=0pt, outer sep=0pt, draw=black, fill=white},
  net sqnode/.style = {square, minimum width=2*\ab cm, inner sep=0pt, outer sep=0pt, draw=black, fill=white},
  net root/.style = {cloud, cloud puffs=15.7, cloud ignores aspect, minimum width=10*\ab cm, minimum height=4*\ab cm, align= center, draw=black, line width =1pt},
  net connect/.style = {line width=1pt, draw=black},
}
\tikzstyle{net node} = [circle,draw, fill=blue!20,inner sep=1.2pt]
\tikzstyle{dec} = [draw, fill=red!20]
\tikzstyle{block} = [draw, rectangular, fill=yellow!20,text centered,rounded
corners]

\definecolor{mypink}{RGB}{234, 175, 169}
\definecolor{myorange}{RGB}{233, 215, 192}

\ifarxiv
\else
\endlocaldefs
\fi

\begin{document}

\ifarxiv
\else
\begin{frontmatter}
\fi

\title{\Bagel: A Fast Bayesian Online Changepoint Detection Algorithm for Linear Models via Posterior Merging}

\ifarxiv
\author{
    Ziyang Yang\thanks{STOR-i Centre for Doctoral Training, Lancaster University}
    \and
    Paul Fearnhead\thanks{Department of Mathematics and Statistics, Lancaster University}
    \and
    Dan Grose\footnotemark[2]
}
\maketitle
\fi

\begin{abstract}
Bayesian online changepoint detection provides a natural framework for quantifying uncertainty about whether and when a change has occurred. It also enables the use of prior information which can increase detection power. Exact Bayesian inference sequentially updates the posterior of whether and when a change has occurred. We present a unified approach to calculating and updating this posterior for a class of regression models covering standard changepoint scenarios, such as changes in mean and slope, whilst also allowing common structure such as seasonal effects. However, naive implementation has a computational cost that increases linearly over time. We, further, introduce approximate updates that keep this cost constant. These are based on novel ideas of merging changepoint times, and are an alternative to existing pruning approaches. We show that merging enables better approximations of the posterior of when a change occurs. Furthermore, we introduce a novel condition for merging that accounts for both the posterior probability of the changepoint times and the similarity of their corresponding post-change parameters. We show that this is more accurate than conditions that only consider the former. We demonstrate the advantage of our approach on simulated data, and the Azure LLM inference traces.
\end{abstract}

\ifarxiv

\else
\begin{keyword}[class=MSC]
\kwdgroup[type=primary]{\kwd{62F15}}
\kwdgroup[type=secondary]{\kwd{62L10}
\kwd{62-08}}
\end{keyword}

\begin{keyword}
\kwd{Anomaly detection}
\kwd{Bayesian online changepoint detection}
\kwd{Bayesian linear models}
\kwd{Fixed-budget approximation}
\kwd{Posterior merging}
\kwd{Real-time system monitoring}
\end{keyword}
\end{frontmatter}
\fi

\section{Introduction}

Detecting structural changes in time-series is important across many areas such as climate change \citep{lund2023good}, environmental science \citep{harris2022scalable}, finance \citep{banerjee2020change}, bioinformatics \citep{jewell2020fast}, medical imaging \citep{stoehr2021detecting}, software reliability \citep{ke2020software} and physics \citep{ward2022poisson}, amongst many others. See \citet{li2026change} for a recent review of the area and applications. Whilst traditionally the focus of methods to detect these changepoints has been on offline analysis, there has been increasing interest in online methods for detecting changepoints due to the importance of detecting changes in real-time, from data such as medical monitoring signals \citep{oneline-dynamic}, telecommunications network data \citep{romano2023log}, and industrial sensor measurements \citep{yang2026mixfocus}.

Frequentist approaches to online changepoint detection are well established and include approaches based on sequential likelihood ratio tests \citep{page1954continuous,page1955test,yu2020note,romano2021fast} and moving window approaches \citep{kirch2018modified}. However, for some applications there are advantages of using a Bayesian approach. Such methods can incorporate prior information about the type and size of change, and they directly quantify uncertainty in both the presence and location of a change. For offline analysis, Bayesian changepoint methods allow inference for multiple changes and tend to involve  Markov Chain Monte Carlo \citep{stephens1994bayesian,Gre95,chib1998estimation,bensonadaptivemcmc}. For a certain class of models, exact calculation and simulation from the posterior is possible \citep{barry1993bayesian,fearnhead2006exact,fearnhead2011efficient} through solving a set of recursions.

The latter class of recursive methods can be implemented in an online manner. These apply to models where the likelihood of the data after a change does not depend on the parameters associated with segments before that change. Exact inference has a computational cost that increases linearly with time, so processing the observation at time $T$ has an $O(T)$ cost. This cost can be reduced to $O(1)$ by introducing an appropriate approximation \citep{fearnhead2007line,adams2007bayesian}. 

For online applications, we often only need to detect whether or not there has been a changepoint prior to the current time. This then involves only considering  models for the data with no or just one change. In this paper we consider such a setting. We specifically consider models where the data can be written as a unknown signal plus noise, and where a change corresponds to a change in the structure of the signal. It is straightforward to construct a general recursion for calculating the posterior in such a model. Importantly, as we are only allowing at most one change, this recursion can be applied to a wider range of models than those of \citet{fearnhead2007line} and \citet{adams2007bayesian} as the data after the change can depend on the parameters associated with the pre-change data. This allows us to, for example, detect changes in slope and changes in mean in the presence of seasonal effects.

Solving our recursion exactly leads to a computational cost per iteration that increases linearly with the amount of data. To avoid this we could use the pruning (or resampling) ideas from \citet{fearnhead2007line} and \citet{adams2007bayesian}: the increasing computational cost comes from having to consider all possible changepoint locations prior to the current time, and pruning involves ignoring some of these possibilities which have lowest probability. We instead propose an alternative, novel approach to reducing the computational cost. This involves merging rather than pruning: we partition the set of possible changepoint locations into a set of consecutive intervals, and then jointly update the posterior probabilities associated with values within an interval. This has the advantage over pruning in that it retains an approximation to the posterior distribution of a change at all time-points, and we show that this leads to a more accurate approximation of the posterior of the change location. 

Merging has previously been suggested, for a different class of changepoint models, by \citet{shamp2021computationally}, though without the idea of approximating the posterior distribution on the time of a change that motivates our use. Our implementation of how we merge is also different. Merging involves two steps: deciding which two current intervals of changepoint times to merge, and then choosing an approximate posterior for the parameters of the model given a change occurring within the merged interval. The approximation of merging comes from assuming the same posterior for the parameters for each changepoint time in the interval -- as the true posteriors will differ for different changepoints. \citet{shamp2021computationally} choose to merge the two intervals with the lowest current posterior probability, and then approximate the posterior for this merged interval based on minimising some distance to the posteriors that were associated with the two intervals that have been merged. We instead use a criterion that aims to minimise the difference between the posteriors on the post-change parameters prior to and after merging. Importantly, we do not just merge intervals based on their current posterior probability, but also on how similar the posteriors associated with each interval are. 

The intuition behind our merging criteria is that the future updates of the posterior probabilities associated with each changepoint location depend on the posterior for the post-change parameter given a change at that location. By merging intervals with similar posteriors for the post-change parameter, we ensure that future approximations to the posterior on the changepoint location are more accurate. We demonstrate empirically that this is the case, and that often we can get substantially more accurate results with our merging criteria than if we just merged based on the posterior probabilities associated with a change in each interval. We name our approach Bagel, from BAyesian online chanGEpoint detection for Linear models.

The outline of the paper is as follows. In Section 2, we define the Bayesian real-time changepoint detection problem and introduce our proposed algorithm. Sections 3, and 4 show the performance of our algorithm on simulated data and real datasets. More examples, proofs, and detailed simulation results can be found in Supplement A. Data and reproducible code can be found in Supplement B. 

Throughout this paper, all vectors are assumed to be column vectors unless explicitly stated otherwise. We denote by $1:t$ the column vector $(1,\cdots,t)^\top$, $1_{1:t}$ and $0_{1:t}$ are  column vectors of length $t$ of  ones and zeros respectively, and $y_{1:t}$ represents the column vector $(y_1,\cdots,y_t)^\top$. The probability density function (pdf) of the normal distribution with mean $\mu$ and variance $\sigma^2$ is denoted by $ N(x ;\mu,\sigma^2) $, the pdf of the Student-t distribution with $\nu$ degrees of freedom, location $l$, and scale $e$ is denoted by $ t_\nu(x;l, e) $, and the pdf of the Inverse-Gamma distribution with shape $\nu$ and scale $\iota$ is denoted by $ IG(x ; \nu,\iota)$. For a matrix $A$, we use $A_{i:j \times k:l}$ to denote the sub-matrix formed by rows $i$ to $j$ and columns $k$ to $l$.

\section{Univariate real-time Bayesian changepoint detection}
\subsection{The changepoint problem}\label{change_prob}
\subsubsection{The model}
Assume an independent data stream $y_1,y_2,…, y_t$ is observed in real-time, we wish to detect the presence, or not, of a changepoint at each time step. More precisely, at a given time $t$, we wish to detect whether there has been a change prior to $t$ under an appropriate model for the data. We will consider a class of linear models that encompasses the standard change-in-mean and change-in-slope problem, as well as more complex models, for example incorporating seasonality. \\

Let $\tau$ be the unknown time of the changepoint, where $\tau=\infty$ denotes no change. The observed data $y_t$ at any time $t$ is modelled as:
\[y_t = a_t^\top \beta + b_{t,\tau}^\top \gamma + \sigma \epsilon_t.
\]
Here $a_t$ and $b_{t,\tau}$ are known $d_1$ and $d_2$ dimensional vectors, while $\beta$ and $\gamma$ are unknown $d_1$ and $d_2$ dimensional parameters. The term $a_t^\top \beta$ gives the mean of the model if there has been no change prior to time $t$, while $b_{t,\tau}^\top \gamma$ specifies the change in this mean due to a changepoint. To be consistent with this we have $b^\top_{t,\tau} = 0_{1:d_2}$ if $\tau\geq t$.  Finally, $\epsilon_t$, is the realisation of a standard Gaussian random variable, and $\sigma>0$ is the standard deviation of the noise in the model.

Consequently, at any time $t$, the model for the data up to time $t$, $y_{1:t}$, can be written as:
\[
y_{1:t} = A_{1:t} \beta + B_{1:t,\tau}\gamma + \sigma \epsilon_{1:t},
\]
where $\epsilon_{1:t}$ is a column vector of i.i.d Gaussian $N(0,1)$ realisations, and the $t$th rows of matrices $A_{1:t}$ and $B_{1:t,\tau}$ are $a_t^\top$ and $b_{t,\tau}^\top$ respectively. If $\tau\geq t$ then $B_{1:t,\tau}$ is the zero matrix. Since $\tau$ is unknown, at each time $t$ we consider $t$ potential models: one corresponding to no change and others where change occurs at $\tau = 1, 2,...,t-1$. Our task is to determine whether there is significant evidence supporting the presence of a change and to infer the posterior distribution of the location of any change. 

\subsubsection{The prior for the model parameters}\label{sec:prior}
The likelihood of the observation depends on the parameters $(\beta,\gamma)$, if the variance of the observations is known, or $(\beta,\gamma, \sigma)$ otherwise. We take a Bayesian approach and assume a prior for these parameters. Furthermore, to enable efficient calculation of the posterior for the presence and location of a changepoint we will use conjugate priors. 

For our model, the conjugate priors are Gaussian for $(\beta,\gamma)$ conditional on $\sigma$, with, if it is unknown, an independent inverse gamma prior for $\sigma$. We will have fixed priors for $\beta$ (and if unknown, $\sigma$), but we will allow the conditional prior for $\gamma$ to potentially depend on both $\tau$ and $\beta$. Specifically, the joint prior for $(\beta,\gamma)$ given $\sigma$ and $\tau$ is
\[\beta,\gamma|\sigma^2,\tau \sim N(\mu^{\tau},\sigma^2 \Sigma^{\tau})\]
with prior mean $\mu^{\tau} \in \mathbb{R}^{d_1+d_2}$ and prior scaled covariance matrix $\Sigma^{\tau}\in \mathbb{R}^{(d_1+d_2)\times(d_1+d_2)}$. 
The constraint of a fixed prior for $\beta$ given $\sigma$ means that $\mu^{\tau}_{1:d_1}$ and $\Sigma^\tau_{1:d_1 \times 1:d_1}$, that is the marginal prior mean and variance of $\beta$, are constant as we vary $\tau$. If $\sigma$ is unknown, the prior is given by:
\[\sigma^2 \sim IG(\nu,\iota).\]

While it is often possible to use improper priors for $\beta$ and $\sigma$ - the conditional prior for $\gamma$ must be proper in order to have a well-defined posterior probability for whether there is a changepoint prior to any given time $t$. The choice of the prior for $\gamma$ will affect the power of detecting different sizes and types of change.

Here we give three examples to illustrate our model framework (with derivations of the stated priors given in 
\ifarxiv
Appendix \ref{Supplement A:prior}\fi). In each case we present the priors for the case where the noise variance is known, but extending them to the unknown variance case only requires specifying the inverse Gamma prior for the variance. Through we denote user-chosen constants in the prior mean by $\mu_i$, for $i=1,2,\dots$ and in the variance by $\delta_i>0$ for $i=1,2,\dots$.

\noindent
\textbf{Example 1 (Change-in-mean model):}
We first consider the change-in-mean model {\color{black}where $d_1=d_2=1$}. The model can be written as:
\[A_{1:t}=1_{1:t}, B_{1:t,\tau}=\left[\begin{array}{c}
       0_{1:\tau} \\
       1_{\tau+1:t} 
    \end{array} \right]
\]
Here $\beta$ is the mean before the change and $\gamma$ represents the shift in the mean after the changepoint. There are two natural choices for the prior distribution in this setting: one where the pre-change mean and the change in mean are independent, and another where the pre-change and post-change means are independent and identically distributed. 

For the former case we have $
\mu = \left[\begin{array}{c}
       \mu_1 \\
       0 
    \end{array} \right]$ and $\Sigma = \left[\begin{array}{cc}
      \delta_1 & 0  \\
      0& \delta_2 
    \end{array} \right]$.  For the latter case, we have the same prior mean, but the covariance structure changes to
$
\Sigma = \begin{bmatrix}
    \delta_1 & -\delta_1  \\
    -\delta_1 & 2\delta_1  
\end{bmatrix}.
$

\noindent
\textbf{Example 2 (Change-in-slope models):}
Next, we consider the change-in-slope {\color{black}where $d_1=d_2=2$}. The model can be written as:
\[A_{1:t}=[1_{1:t} ~~~1:t], B_{1:t,\tau} = \left[\begin{array}{cc}
       0_{1:\tau}  & 0_{1:\tau} \\
       1_{(\tau+1):t} & (\tau+1):t \\
    \end{array} \right] \] 
Here $\beta$ defines the linear trend before the change, and $\gamma$ specifies the shift in the intercept and slope at the changepoint.

\begin{figure}
    \centering
\includegraphics[width=0.7\linewidth]{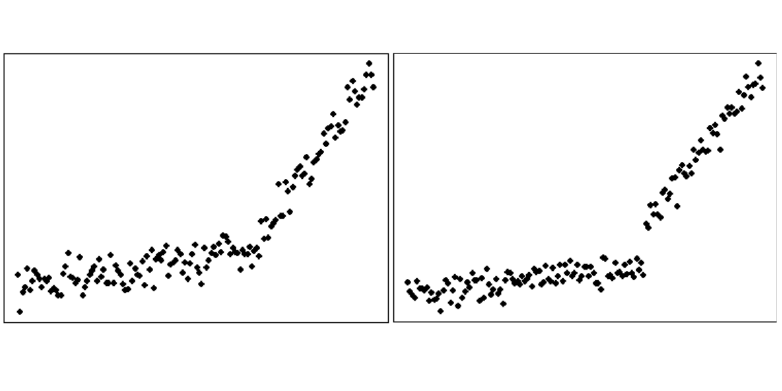}
\caption{An example of continuous change (left) and discontinuous change (right) in a linear trend model.}
    \label{fig:example-slope}
\end{figure}

There are two natural types of change we can allow in this model, depending on whether or not we require continuity of the trend line at the changepoint, as illustrated in Figure \ref{fig:example-slope}. If we wish to detect a continuous change, also known as a change-in-slope \citep{fearnhead2011efficient}, then a natural model is that the change-in-slope, $\gamma_2$, has mean 0 and is independent of the pre-change trend. The distribution of $\gamma_1$ conditional on $\gamma_2$ is then deterministic due to the continuity constraint. This constraint implies for a change at $\tau$ that the trend is continuous at $\tau$: $\gamma_1+\tau\gamma_2=0$. This gives a prior for $(\beta,\gamma)$ that has mean and covariance defined by the parameters:
\[
\mu^{\tau}=(\mu_1,\mu_2,0,0)^\top,\mbox{ and }~~
\Sigma^{\tau} = 
\begin{bmatrix}
\delta_1 & 0 & 0& 0 \\
0 & \delta_2 & 0 &0 \\
0 & 0& \tau^2 \delta_3 & -\tau \delta_3  \\
0 & 0& -\tau \delta_3 & \delta_3 \\    
\end{bmatrix}.
\]
For a discontinuous change 
a natural choice of prior would assume that the distribution of the trend line starting at time 0 is independent of and has the same distribution of the trend line starting at time $\tau$. This requires the value of the mean function at $\tau$, $\beta_1+\gamma_1+\tau(\beta_2+\gamma_2)$ has the same distribution as $\beta_1$, and the new slope $(\beta_2+\gamma_2)$ has the same distribution as $\beta_2$. This implies a prior with
\[
\mu^{\tau}=(\mu_1,\mu_2,-\mu_2\tau,0)^\top,\mbox{ and }~~
\Sigma^{\tau} = \left[
\begin{array}{cccc}
\delta_1 & 0 & -\delta_1& 0 \\
0 & \delta_2 & 0 & -\delta_2 \\
-\delta_1 & 0& 2\delta_1+\tau^2\delta_2  &  -\tau\delta_2 \\
0 &  -\delta_2 & -\tau\delta_2 & 2\delta_2  \\
\end{array}
\right].
\]

\noindent
\textbf{Example 3 (Change-in-seasonality model):}
The framework can include seasonality. For simplicity, we consider the seasonal model with one Fourier-harmonic seasonality period $P$ and known variance. In this case $d_1=d_2=2$, and the model can be written as:
\[A_{1:t} = \left[\sin \frac{2\pi(1:t)}{P}, ~~ \cos \frac{2\pi(1:t)}{P} \right],  B_{1:t,\tau} = \left[\begin{array}{cc}
       0_{1:\tau}& 0_{1:\tau} \\
       \sin \frac{2\pi((\tau+1):t)}{P} & \cos \frac{2\pi((\tau+1):t)}{P}\\
    \end{array} \right]. \]
Here $\beta$ defines the seasonal pattern before the change, and $\gamma$ specifies the shift in the strength of each Fourier-harmonic profile after the changepoint.

A natural choice of prior would be for the distribution of the Fourier coefficients before and after the change to be independent. In this case the mean will be  $\mu^T = \left[\mu_{\sin},\mu_{\cos},0,0\right]$, and if the marginal variance of $(\beta_1,\beta_2)^T$ is $\Sigma_s=\left[\begin{array}{cc}
      \delta_{\sin} & 0   \\
      0& \delta_{\cos} 
    \end{array} \right]$, then the prior variance will be  $\Sigma^\tau = \left[\begin{array}{cc}
      \Sigma_s & -\Sigma_s  \\
      -\Sigma_s& 2\Sigma_s \\
    \end{array} \right]$.

\subsubsection{Prior for the changepoint} \label{prior_for_change}
To complete our model specification we need a prior for $\tau$. At time $t$, denote the prior probabilities by $\Pr_t(\tau=i)$, for $i=1,\ldots,t-1$ and $\Pr_t(\tau\geq t)$. The latter probability will encompass the event that a change has not occurred by time $t$.

One approach to defining the prior at time $t$ is to construct a prior for $\tau$ taking values in $1,\ldots$ and $\tau=\infty$ (i.e. no change), and then define $\Pr_t(\tau)$ to be consistent with this. However, this is not necessarily appropriate in an online setting where we are monitoring a data stream and will intervene if we detect a change. In this case, there is information in the fact that we are still monitoring the data stream at time $t$ that should affect our prior for $\tau$. Intuitively, this information would mean that it is less likely for $\tau$ to take smaller values, for which there would be more power for detecting a change by $t$.

Thus we suggest the following approach. First let us define the Bayes factor for a change at $\tau=i$ against no change $\tau\geq t$ as
\[
BF_t(i) = \frac{p(y_{1:t}|\tau=i)}{p(y_{1:t}|\tau\geq t)},
\]
where the terms on the right-hand side are marginal likelihoods, defined as
\[
p(y_{1:t}|\tau) = {\color{black}\int} \int p(y_{1:t}|\tau,\beta,\gamma,\sigma) p(\beta,\gamma|\tau,\sigma)  \mbox{d}\beta \mbox{d}\gamma ,
\]
in the case where $\sigma$ is known, and where we would further integrate out with respect to the prior for $\sigma$ if it were unknown. The Bayes factor for a change prior to $t$ is obtained by averaging these Bayes factors with respect to the conditional prior for $\tau$ assuming a change by time $t$,
\[
BF_t = \sum_{i=1}^{t-1} BF_t(i)  \Pr_t (\tau=i|\tau < t).
\]
So the posterior probability of a change before time $t$ is
\[
p_t(\tau < t| y_{1:t}) = \frac{(1-\Pr_t(\tau \geq t) ) BF_t }{\Pr_t(\tau \geq t) + {\color{black}(1-\Pr_t(\tau \geq t))} BF_t}.
\]
If we use a monitoring scheme which detects a change if $p_t(\tau < t| y_{1:t}) \geq c_t$ for some constant $c_t$, this will be equivalent to detecting a change based on the Bayes Factor being greater than a constant, i.e.
\[
BF_t \geq \frac{c_t \Pr_t(\tau\geq t)}{(1-c_t)(1-\Pr_t(\tau\geq t))}.
\]
In the absence of specific prior information, it would be natural to detect a change based on the Bayes Factor being above a threshold that is constant over time. This is easiest to implement with $c_t$ and $\Pr_t(\tau\geq t)$ both being constant. Furthermore, if we take these to be constant, and tune the threshold for detecting a change based on properties of the test under data where there is no change, then this would make our test invariant to the choice of prior probability $\Pr_t(\tau\geq t)$: as choosing a different prior probability would lead to tuning a different threshold such that we were implementing exactly the same test. We take this approach in our simulation study. 

The test does depend on our prior for $\tau$ conditional on $\tau < t$. It is natural to define this prior as
\[
\Pr_t(\tau=t-k | \tau < t) \propto p^k, \mbox{ for} \quad k= 1,\ldots,t-1.
\]
This is a (truncated geometric) prior for the time since the change. {\color{black}This prior is time dependent, and is updated sequentially as new observations arrive, ensuring that $\Pr_t(\tau < t)$ remains constant while the probabilities of having a change at each location are adjusted.} A specific case, which we use in our simulations, is when $p=1$ and we have a uniform distribution for this conditional distribution of the change location. In this case the Bayes Factor is similar to the Shiryaev-Roberts test statistics which is known to have good properties \citep{polunchenko2010optimality}.

\subsection{Sequential Updating}

Recall that we wish to calculate the posterior probability of a change, its location and the parameters of the model at each time $t$. To do this in a computationally efficient way, we need to update these posterior probabilities and distributions sequentially from time $t-1$ to time $  t$ given $y_t$.

The posterior probability for $\tau$ at time $t$ satisfies
\[
\Pr_t(\tau|y_{1:t}) \propto \Pr_t(\tau) p(y_{1:t}|\tau). 
\]
We will derive a recursion for the right-hand side. To do this at time $t$ we define a set of weights: Let $w_{i,t}=\Pr_t(\tau=i) p(y_{1:t}|\tau=i)$, for $i=1,\ldots,t-1$, and define $w_{0,t}=\Pr_t(\tau\geq t) p(y_{1:t}|\tau \geq t)$. We are subscripting the weights with the pair $i,t$, where $t$ denotes the current time and $i$ the value of the changepoint, $\tau$, but with slight abuse of this notation we have $i=0$ to denote no changepoint prior to $t$.

The following result gives the update for these weights in terms of the predictive density for $y_t$ given $y_{1:t-1}$.
\begin{theorem}\label{Thm:weight_updating}
For $t>1$,
\begin{eqnarray*}
w_{i,t} &=& w_{i,t-1} \frac{\Pr_t(\tau=i )}{\Pr_{t-1}(\tau=i)}p(y_t|y_{1:t-1},\tau=i) \mbox{ for }  i=1,\ldots, t-2, \mbox{and}~~ t>2 \\
 w_{0,t} & = & {\color{black}w_{0,t-1} p(y_t|y_{1:t-1},\tau\geq t), }\mbox { and}\\
w_{t-1,t} & = & w_{0,t-1} \frac{\Pr_t(\tau=t-1)}{Pr_{t-1}(\tau\geq t-1)} p(y_t |y_{1:t-1},\tau=t-1).
\end{eqnarray*}
With the weights initialised at $t=1$ by
\begin{eqnarray*}
w_{0,1} & = \Pr_t(\tau\geq 1)p(y_{1}|\tau\geq 1).
\end{eqnarray*}
\begin{proof}
\ifarxiv
See Appendix \ref{sequential_update}
\fi
\end{proof}
\end{theorem}

We can implement these recursions by recursively calculating the posterior of the model parameters, and hence the predictive distribution, for each possible value of $\tau$. The following result gives updates for the posterior for the parameters in the case $\sigma$ is unknown, but these can be applied in the case $\sigma$ is known by ignoring the update for the posterior of $\sigma$. 

As the prior for the post-change parameter can depend on the time of change, in the case of no changepoint we will update only the marginal posterior for $\beta$. To introduce notation, at time $t-1$ let the posterior for $\beta$ and $\sigma$ given there is no change be
\[
p(\beta|y_{1:t-1},\sigma,\tau\geq t-1 ) \sim N \left( \mu^{t-1,0}, \sigma^2\Sigma^{t-1,0} \right), \]
\[\mbox{ and }~~
p(\sigma^2|\tau \geq t-1) \sim IG ( \nu^{t-1,0},\iota^{t-1,0} ),
\]
where $ \mu^{t-1,0}$ is a $d_1$ dimensional vector and $\Sigma^{t-1,0}$ is a $d_1\times d_1$ matrix. Similarly, we denote the posterior given a change at $i=1,\ldots,t-2$ as
\[
p(\beta,\gamma|y_{1:t-1},\sigma,\tau=i ) \sim N \left( \mu^{t-1,i}, \sigma^2\Sigma^{t-1,i} \right),\]\[ \mbox{ and }~~
p(\sigma^2|\tau =i) \sim IG ( \nu^{t-1,i},\iota^{t-1,i} ),
\]
where now $ \mu^{t-1,i}$ is a $d_1+d_2$ dimensional vector and $\Sigma^{t-1,i}$ is a $(d_1+d_2)\times(d_1+d_2)$ matrix. Finally, we will need the posterior distribution for $(\beta,\gamma)$ and $\sigma$ assuming a change at time $t-1$. Using the same notation as above, this can be obtained from the posterior for $\beta$ and $\sigma$ given no change prior to $t-1$.
\begin{theorem} \label{Thm:t-1}
Let the prior mean for a change at $t-1$ be $\mu^{t-1}=[\mu_\beta^\top,\mu_\gamma^\top]^\top$, and the prior covariance be
\[
\sigma^2\Sigma^{t-1} = \sigma^2\begin{bmatrix}
    \Sigma_{\beta,\beta} & \Sigma_{\beta,\gamma} \\
    \Sigma_{\gamma,\beta} & \Sigma_{\gamma,\gamma}
\end{bmatrix},
\]
so that $\mu_{\beta}$ denotes the prior mean of $\beta$, $\Sigma_{\beta,\beta}$ the prior variance of $\beta$, $\Sigma_{\beta,\gamma}$ the prior covariance between $\beta $ and $\gamma$, and so on. Then the distribution of $(\beta,\gamma,\sigma)$ given $y_{1:t-1}$ and a change at $\tau=t-1$ has parameters
\[
\mu^{t-1,t-1}=\left[ \begin{array}{c} \mu^{t-1,0} \\ \mu_{\gamma} + \Sigma_{\gamma,\beta}\Sigma_{\beta,\beta}^{-1}(\mu^{t-1,0}-\mu_{\beta})\end{array}\right],
\]
\[
\Sigma^{t-1,t-1}=\begin{bmatrix}
\Sigma^{t-1,0}  & \Sigma^{t-1,0}\Sigma_{\beta,\beta}^{-1}\Sigma_{\beta,\gamma} \\ \Sigma_{\gamma,\beta}\Sigma_{\beta,\beta}^{-1}\Sigma^{t-1,0}
& \Sigma_{\gamma,\gamma}+\Sigma_{\gamma,\beta}\Sigma_{\beta,\beta}^{-1} (\Sigma^{t-1,0}-\Sigma_{\beta,\beta} ) \Sigma_{\beta,\beta}^{-1}\Sigma_{\beta,\gamma}
\end{bmatrix}
\]
with $\nu^{t-1,t-1}=\nu^{t-1,0}$ and $\iota^{t-1,t-1}=\iota^{t-1,0}$.
\end{theorem}
\begin{proof}
\ifarxiv
See Appendix \ref{sequential_update}
\fi
\end{proof}

We can update these parameters as follows.
\begin{theorem} \label{thm:posterior}
Fix iteration $t-1$, define $h_i$ as the following feature vector:
\[h_i=
\begin{cases}
  [a_{t-1}^\top,b_{t-1,i}^\top]^\top & 0<i \leq t-2, \\
   a_{t-1} &i =0,
\end{cases}
\]
For $i=0,1,\ldots,t-2$, define $e_i = y_{t-1} - h_i\mu^{t-1,i}$,$Q= h_i^\top\Sigma^{t-1,i}h_i+1$, $A = \Sigma^{t-1,i} h_i/Q$, The parameter updates for existing changepoint models after observing $y_{t-1}$ follow:
\begin{alignat*}{2}
\Sigma^{t,i} &= \Sigma^{t-1,i} -A A^\top Q,  \quad &\mu^{t,i} &= \mu^{t-1,i} + Ae, \\
\nu^{t,i}&=\nu^{t-1,i}+\frac{1}{2}, \quad &\iota^{t,i} &= \iota^{t-1,i}+\frac{1}{2}e_i^2/Q. 
\end{alignat*}
\end{theorem}

Finally, given the posterior for the parameters at time $t-1$, which is also the prior for the models at time $t$, we have the following calculations for the predictive density.
\begin{theorem}\label{thm:predictive}
Using the same notation for the parameters of the posteriors as in Theorem \ref{thm:posterior}, the predictive density can be calculated as 
\[
P(y_t|y_{1:t-1},\tau = i) = N\left(y_t;h_i\mu^{t,i},\sigma^2(1+h_i^\top\Sigma^{t,i}h_i)\right)
\]
if $\sigma$ is known, and
\[
P(y_t|y_{1:t-1},\tau = i) = t_{2\nu^{t,i}}\left(y_t;h_i\mu^{t,i},\frac{\iota^{t,i}}{\nu^{t,i}} \left(1+h_i^\top\Sigma^{t,i} h_i\right)\right)
\]
if $\sigma$ is unknown, where $t_\nu(y_t;l,e)$ is the location-scale t distribution with $\nu$ degrees of freedom, location $l$ and scale parameter $e$.
\end{theorem}

\subsection{Reducing the computational complexity by merging}
\begin{figure}
    \centering
     \subfloat[Simulated data]{ \includegraphics[page=1,width=0.33\linewidth]{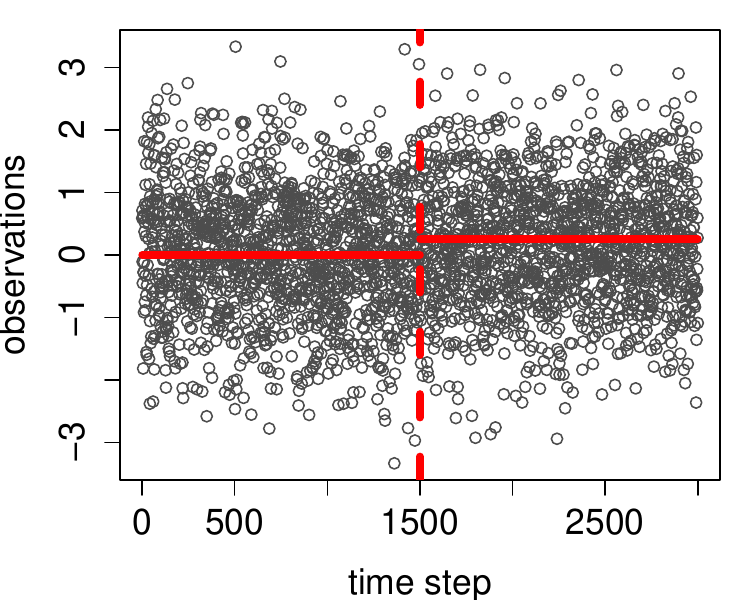}} 
     \subfloat[At step 1600]{\includegraphics[page=2,width=0.33\linewidth]{pic/step_2000_motivation.pdf}}
    \subfloat[At step 2000]{\includegraphics[page=3,width=0.33\linewidth]{pic/step_2000_motivation.pdf}}
    \caption{A failure case example for LP-P. Figure (a) presents the simulated data with a change at 1500. Figures (b) and (c) show the posterior distribution obtained from the exact approach without pruning (grey line). The black dots represent the candidates not pruned by the pruning approach with particle budget $M=50$ at time steps 1600 and 2000, respectively. We can see that candidates with high probability at step 2000 were pruned earlier.}
    \label{fig:common-fail}
\end{figure}

Implementing the univariate Bayesian changepoint detection methods gives a cost per iteration that increases with iteration $t$, so that the overall cost of analysing $T$ data points is quadratic in $T$. To prevent this, we need to approximate the posterior distribution with fewer support points. A common and straightforward approach is to prune all but the most $M$ probable candidates \citep{adams2007bayesian, saatcci2010gaussian}, in which the computational complexity can be reduced to $O(TM)$. For simplicity, we call this approach LP-P (lowest probability pruning). However, such a method does not always work well. It reduces the support of the posterior for the change, which can be important if we wish to calculate credible intervals. Furthermore, if different changepoint values have substantially different posteriors for the post-change model then their relative probabilities can change substantially as we observe new data, as shown in Figure \ref{fig:common-fail}.

\subsubsection{Merging criteria}
To overcome these issues, we first use the idea of merging the posterior distribution for the parameters for different change locations as a way to reduce the computational cost whilst keeping the full support of the posterior. Second, we take account of the similarity of the posterior distributions of the parameters when deciding which distributions to merge. We will look at the similarity of the posteriors in terms of the parameters, or the function of parameters, that affect the distribution of the data after the change. Denote such parameters as $\theta=g(\beta,\gamma)$, where $g(\cdot)$ is a linear transformation that maps the parameters $(\beta, \gamma)$ to a smaller set of parameters, $\theta$, that determine the distribution of the data after the change. For example the post-change mean in Example 1 or the post-change intercept and slope in Example 2.

To explain our approach, it is simplest to first consider what happens when we first merge posteriors. Fix the time, $t$, when this happens, and denote the posterior density function for the post-change parameter as $f_i(\theta)$ when change occurs at $i$. We have $t-1$ possible values for the location of a change, assuming a change has occurred, and these will have weight and posterior distribution denoted by the pair $\{w_{i,t}, f_i(\theta)\}, 1 \leq i \leq t-1$.

Our merge procedure will replace $f_i(\theta)$ with $f_{i+1}(\theta)$ for some appropriately chosen $i$. We will choose $i \in\{1,\ldots,t-2\}$ to be the value for which the total variation distance between the posterior for $\theta$ and the resulting approximation is minimised. Let $TV(h_1(\theta),h_2(\theta))=\frac{1}{2}\int |h_1(\theta)-h_2(\theta)|\mbox{d}\theta$ be the total variation distance between two densities, $h_1(\theta)$ and $h_2(\theta)$. Then if the true posterior is $h_1(\theta)=\sum_{i=1}^{t-1} w_{i,t} f_{i}(\theta)$, the approximation after we replace $f_i(\theta)$ with $f_{i+1}(\theta)$ is $h_2(\theta)=h_1(\theta)-w_{i,t}(f_i(\theta)-f_{i+1}(\theta))$, and
\begin{eqnarray*}
TV\left(h_1(\theta),h_2(\theta)\right) &=&
\frac{1}{2}\int \left|h_1(\theta)-\left(h_1(\theta) -w_{i,t}(f_i(\theta)-f_{i+1}(\theta)) 
\right)\right|\mbox{d}\theta \\ 
&=& \frac{1}{2} \int \left| w_{i,t}(f_i(\theta)-f_{i+1}(\theta))\right|\mbox{d}\theta \\
&=& w_{i,t} TV(f_i(\theta),f_{i+1}(\theta)).
\end{eqnarray*}
Thus we will choose to merge component $i$ which minimises 
\begin{equation} \label{eq:TV} 
w_{i,t} TV(f_i(\theta),f_{i+1}(\theta)).
\end{equation}

This will maintain the same support for the changepoint locations, but will reduce the computational cost. At any time, if we have $M$ distinct posterior densities for $\theta$, then we only need to update these posteriors when we get a new observation. To update the weights associated with a given posterior, we store these as the sum of the weights for that posterior and the relative weight for each change location. The latter will not change, so we only need to update the former weight.

So at any iteration $t$, after applying merging, we will have consecutive runs of changepoint locations that will have the same posterior for $\theta$. At the current iteration, denote the largest value of such runs by an ordered set $\mathcal{T}=\{j_1,j_2,\ldots,j_M\}$. Let $w_{i,t}$ for $i=0,\ldots,M$ denote the set of weights such that for $i\in\{1,\ldots,M\}$, $w_{i,t}$ is the sum of the weights associated with change locations $\{j_{i-1}+1,\ldots,j_i\}$, with $j_0=0$. Also we store a set of $M$ ratio vectors $\mathcal{R}=\{r^{(1)},\ldots,r^{(M)}\}$ which denote the proportion of the weight $w_i$ associated with each changepoint location in the run from $j_{i-1}+1$ to $j_i$. The vector of weights associated with these candidate locations will be $w_{i}r^{(i)}$, or equivalently the vector of posterior probabilities for change locations would be
\[
(w_0, w_1r^{(1)},\ldots, w_Mr^{(M)}),
\]
with, after normalisation, the first component being the probability of no change, and the remaining components being the probability of a change at $1,\ldots,t-1$. 

The idea is that we can update this representation by just updating the weights, and the parameters associated with each distinct posterior -- which has a computational cost proportional to $M$. At each iteration we will also add a new candidate location for a change, and to keep the number of distinct posteriors as $M$ we will need to merge a pair of distinct posteriors for $\theta$ if $t>M$. As before let $f_i(\theta)$ denote the posterior associated with the $i$th run of changepoint locations. Using the same criteria on minimising the approximation based on the Total Variation distance, we will replace $f_i(\theta)$ with $f_{i+1}(\theta)$ for the value $i\geq1$ which minimises $w_iTV(f_i(\theta),f_{i+1}(\theta)$.  
Algorithm~\ref{algorithm_2} describes one iteration of the algorithm.

\begin{algorithm}[]
\footnotesize
\SetKwInOut{Input}{input}\SetKwInOut{Output}{output}
    \textbf{Input:} existing candidates set $\mathcal{T}=\{0,i_1,\ldots,i_N\}$, {\color{black}set of ratio vectors $\{r^{(1)},\ldots,r^{(N)}\}^*$}, weights $\{w_{i,t-1}\}$ and parameters for the posterior distributions $\Theta_i$ for $i \in \{0,\ldots,N\}$.\\
    \KwData{Observe $y_{t}$ at time $t$}
    \BlankLine
    {\color{purple}Introducing new changepoint candidate}\\
    Calculate $\Theta_{N+1}$, the parameters of the posterior distribution given $\tau=t-1$ using Theorem \ref{Thm:t-1}.\\
    $\mathcal{T}\longleftarrow \mathcal{T}\cup \{t-1\}$ \\
    {\color{purple}Updating weights and posterior distributions}\\
    \For{$i\in \{0,\ldots,N+1\}$}{
    Calculate $w_{i,t}$ based on Theorems \ref{Thm:weight_updating} and \ref{thm:predictive}.} 
    Update parameters of posterior distribution, $\Theta_{0},\ldots,\Theta_{N+1}$ using Theorem \ref{thm:posterior}.\\
    Normalize $w_{i,t}$ for $i \in \{0,\ldots,N+1\}$.\\
    {\color{purple}Decide if there is a changepoint}\\
    \If{$1-w_{0,t}>c$}{
    \Output{Change detected; $\mathcal{T}, {\color{black}\{r^{(1)}, \ldots,r^{(N+1)}\}^*}, \{\Theta_0, \ldots,\Theta_{N+1} \},w_{0:N+1,t} $}}
    {\color{purple} Merging step}\\
    \If{$N+1>M$}{
      Search the index $i$ that gives the minimum error (\ref{eq:TV}) for $i \in \{1,\ldots,N\}$.\\
       {\color{black}*}Update the ratio vector $\{r^{(i+1)}\}\longleftarrow(w_{i,t}+w_{i+1,t})^{-1} ( w_{i,t}r^{(i)},w_{i+1,t}r^{(i+1)})$.\\
      Combine $w_{i+1,t} \longleftarrow w_{i,t}+w_{i+1,t}$.\\
      Remove candidate $i$, $\Theta_i$, $r^{(i)}$ and $w_{i,t}$.\\
      $N\longleftarrow N-1$
    }
    \Output{ $\mathcal{T},  {\color{black}\{r^{(1)}, \ldots,r^{(N+1)}\}^*}, \{\Theta_0, \ldots,\Theta_{N+1} \},w_{0:N+1,t}$ }
    
\caption{one iteration of \Bagel}   
\textbf{Note:} The set marked with $*$ can be omitted, resulting in constant memory complexity.
\label{algorithm_2}
\end{algorithm}\DecMargin{0.1em}

\subsubsection{Calculating the total variation distance}
Now we will show how to calculate the total variation, which is needed for our merging step. This can be calculated exactly if $\theta$ is 1-dimensional, but we need to resort to an approximation in higher dimensions.

\noindent
\textbf{One dimensional parameter}\\
Calculating the total variation between two univariate distributions, such as in the change-in-mean case with known variance, is straightforward. In this case, $\theta$ will have a univariate Gaussian distribution, and we use the following result for the total variation distance between two univariate Gaussians.
\begin{prop}The total variation between two univariate Gaussian distributions with means $\mu_i$ and $\mu_{i+1}$, and variances $\sigma^2\Sigma_i$ and $\sigma^2\Sigma_{i+1}$ respectively is: 
\begin{align*}
     \left[\Phi (\frac{c_1 - \mu_{i+1}}{\sqrt{\sigma^2\Sigma_{i+1}}}) - \Phi (\frac{c_2 - \mu_{i+1}}{\sqrt{\sigma^2\Sigma_{i+1}}}) + \Phi (\frac{c_2 - \mu_i}{\sqrt{\sigma^2\Sigma_i}}) - \Phi (\frac{c_1 - \mu_i}{\sqrt{\sigma^2\Sigma_i}})\right].
\end{align*}
where $(c_1, c_2) = \frac{\mu_i \sigma^2\Sigma_{i+1} - \mu_{i+1} \sigma^2\Sigma_i \mp \sqrt{\sigma^2\Sigma_i\sigma^2\Sigma_{i+1}}\sqrt{(\mu_i - \mu_{i+1})^2 +(\sigma^2\Sigma_{i+1} - \sigma^2\Sigma_i)\log \frac{\sigma^2\Sigma_{i+1}}{\sigma^2\Sigma_i} }}{\sigma^2\Sigma_{i+1} - \sigma^2\Sigma_i}$ and $\Phi(c)$ is the Gaussian CDF.
\end{prop}
\begin{proof}
\ifarxiv
See Appendix \ref{Supplement A:gaussian}
\fi
\end{proof}
\noindent
\textbf{Multi-dimensional parameters}\\
 In higher dimensions, it is often intractable to find the exact closed form for the total variation between two distributions. Instead, we use the following tractable bound as an approximation to the total variation distance.
 
 For any pair of distributions for distributions $\theta$ over $R^d$ with densities $f_i(\cdot)$ and $f_{i+1}(\cdot)$, Pinsker’s inequality \citep{pinsker1964information} states that 
\[TV(f_i(\theta), f_{i+1}(\theta)) \leq \frac{1}{\sqrt{2}}\sqrt{\mathrm{KL} (f_i(\theta) \parallel f_{i+1}(\theta))} = \frac{1}{\sqrt{2}}\sqrt{\int f_i(\theta)\log\frac{f_i(\theta)}{f_{i+1}(\theta)}d\theta},\]
where KL denotes the Kullback–Leibler divergence. The KL divergence for our posteriors, which will be either multivariate Gaussian or Normal Inverse Gamma, is given by the following results.

\begin{lemma}\citep{williams2006gaussian}
    The KL between two multivariate Gaussian distributions with means $\mu_i$ and $\mu_{i+1}$, and variances $\sigma^2\Sigma_i$ and $\sigma^2\Sigma_{i+1}$ respectively, is:
    \begin{align*}
        D_{KL}&(N(\mu_i, \sigma^2\Sigma_i)||N(\mu_{i+1}, \sigma^2\Sigma_{i+1})) = \\
        &\frac{1}{2}\left(tr(\Sigma_{i+1}^{-1}\Sigma_i-I)+\frac{1}{\sigma^2}(\mu_{i+1} - \mu_i)^\top\Sigma_{i+1}^{-1}(\mu_{i+1} - \mu_i)+\log\det(\Sigma_i^{-1}\Sigma_{i+1})\right)
    \end{align*}    
\end{lemma}

\begin{lemma}\label{KL_normal_gamma}\citep{soch2016kullback} 
     The KL distance between two normal inverse gamma distributions with means $\mu_i$ and $\mu_{i+1}$, scaling variance $\Sigma_i$ and $\Sigma_{i+1}$, shape $\nu_i$ and $\nu_{i+1}$, scale $\iota_i$ and $\iota_{i+1}$ respectively, is
         \begin{align*}
         D_{KL}[NIG&(\mu_i,\Sigma_i,\nu_i,\iota_i)||NIG(\mu_{i+1},\Sigma_{i+1},\nu_{i+1},\iota_{i+1})]=\\
        &\quad \frac{1}{2}\left[(\mu_{i+1}-\mu_i)^\top\Sigma_{i+1}^{-1}(\mu_{i+1}-\mu_i)\frac{\nu_i}{\iota_i} + tr(\Sigma_{i+1}^{-1}\Sigma_i-I)-
        {\color{black}\log\frac{\det(\Sigma_i)}{\det(\Sigma_{i+1})}}\right]\\
        &\quad +\nu_{i+1} \log\frac{\iota_i}{\iota_{i+1}} - \log \frac{\Gamma(\nu_i)}{\Gamma(\nu_{i+1})}+(\nu_i-\nu_{i+1})\psi(\nu_i)-(\iota_i- \iota_{i+1})\frac{\nu_i}{\iota_i},
    \end{align*}
    where $\Gamma(\cdot)$ is the gamma function and $\psi(\cdot)$ is the digamma function.
\end{lemma}

Our algorithm, Bagel, implements merging with this criteria, based on the TV-distance or its approximation, to choose which pair of changepoint sets to merge. To understand the separate benefits of merging rather than pruning from the choice of criteria to merge, in the following we will also compare Bagel to an algorithm that merges the set of changepoints with lowest weight. We call this latter method LP-M. Thus, comparing LP-M with LP-P indicates any advantage of merging over pruning and comparing Bagel with LP-M shows any advantage of our new merging criteria.

\begin{figure}[]
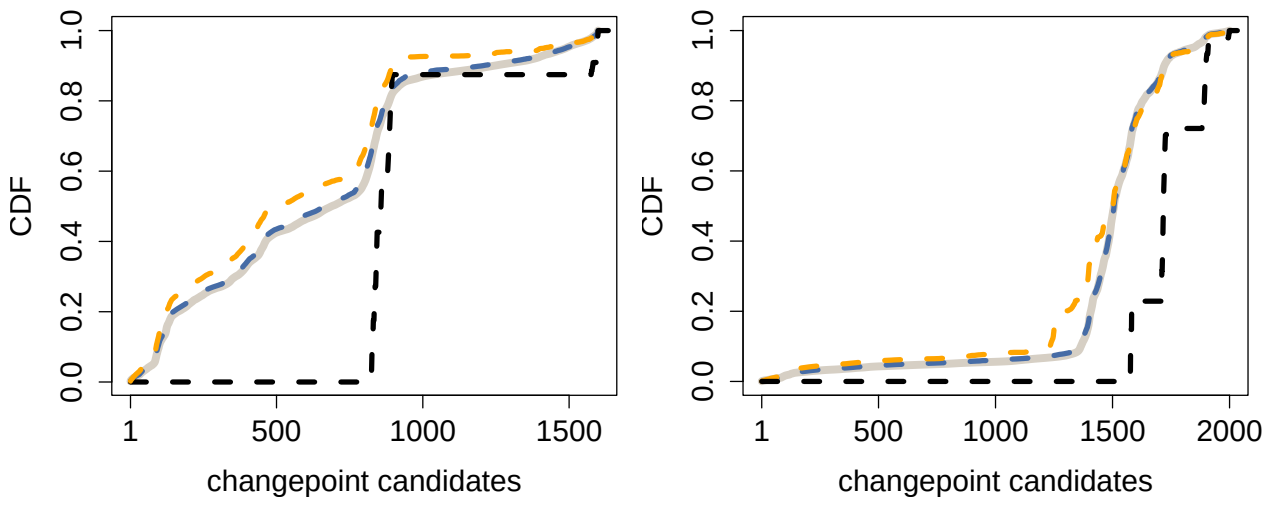

    \centering
     \subfloat[At step 1600]{ \includegraphics[page=4,width=0.33\linewidth]{pic/step_2000_motivation.pdf}} 
     \subfloat[At step 2000]{\includegraphics[page=5,width=0.33\linewidth]{pic/step_2000_motivation.pdf}}
    \caption{The CDFs of the exact approach (grey), \Bagel with recovered posterior distribution (blue), LP-M (orange) and LP-P (black) when $M=50$ on the failure example.}
    \label{fig:common-succ}
\end{figure}

We now revisit the example in Figure \ref{fig:common-fail}. Figure \ref{fig:common-succ} presents the cumulative density function (CDF) of the exact approach, Bagel, LP-P and LP-M. It shows that merging preserves a better posterior approximation than pruning, and the total-variation-based merge criterion improves approximation accuracy.

\section{Simulation Results}\label{sec:sim}
We conduct simulations to compare the posterior approximation accuracy of Bagel, LP-P and LP-M to the exact posterior, and examine whether Bagel accurately approximates the exact Bayesian approach at reduced computational cost.

\subsection{Accuracy of the posterior approximation}\label{calibration}
To evaluate posterior approximation accuracy, we first compare the posteriors of changepoint location from Bagel, LP-M, and LP-P to the exact approach on identical streams under identical priors. For each example, we generate $T=1000$ data points and the changepoint location is drawn from the uniform distribution $\tau \sim \mathrm{U}\{200, \dots, 800\}$. The data-generating process shares the same parameters as the priors. We consider three particle budgets $M=50,100,200$ and two priors. Both priors have 0 mean, but the diffuse one has ten times the variance of the main prior. The detailed prior settings and data-generating process are in \ifarxiv Appendix \ref{app:calibration_detail} \fi.

To assess the similarity between the exact approach and the approximate approaches, we measure the average total variation distance over the posterior distribution
\[
\frac{1}{2R}\sum_{r}^R\left[
  \left|\Pr_m^r(\tau \ge T|y_{1:T}) - \Pr_\infty^r(\tau \ge T|y_{1:T})\right|
  + \sum_{i=1}^{T-1} \left|\Pr_m^r(\tau = i|y_{1:T}) - \Pr_\infty^r(\tau = i|y_{1:T})\right|
\right],
\]
and the average Wasserstein distance on the conditional posterior of changepoint location \[
\frac{1}{R}\sum_{r=1}^R\sum_{i=1}^{T-1}
  \left|\Pr_m^r(\tau \le i \mid \tau < T,y_{1:T}) - \Pr_\infty^r(\tau \le i \mid \tau < T,y_{1:T})\right|
\] over $R$ replications, where $\Pr_m^r = \Pr(\tau|y_{1:T}, M=m)$ is the posterior of approximated method with $m$ particles and $\Pr_\infty^r= \Pr(\tau|y_{1:T},M=\infty)$ is the posterior of exact approach at $r$th replicate. 

\begin{table}
\centering
\small
\scalebox{0.75}{%
\begin{tabular}{l|l|r@{\hspace{8pt}}r@{\hspace{8pt}}r|r@{\hspace{8pt}}r@{\hspace{8pt}}r|r@{\hspace{8pt}}r@{\hspace{8pt}}r|r@{\hspace{8pt}}r@{\hspace{8pt}}r}
\hline
& & \multicolumn{6}{c|}{$\mathrm{TV}_{\mathrm{full}}$} & \multicolumn{6}{c}{$W_{1,\mathrm{change}}$} \\\cline{3-14}
& & \multicolumn{3}{c|}{Main} & \multicolumn{3}{c|}{Diffuse} & \multicolumn{3}{c|}{Main} & \multicolumn{3}{c}{Diffuse} \\\cline{3-14}
Example & $M$ & Bagel & LP-M & LP-P & Bagel & LP-M & LP-P & Bagel & LP-M & LP-P & Bagel & LP-M & LP-P \\
\hline
\multirow{3}{*}{\shortstack[l]{Mean change \\ known variance}} & 50 & \textbf{0.032} & 0.072 & 0.218 & \textbf{0.028} & 0.091 & 0.098 & \textbf{2.27} & 5.88 & 103.85 & \textbf{1.48} & 4.27 & 34.65 \\
 & 100 & \textbf{0.015} & 0.042 & 0.130 & \textbf{0.008} & 0.039 & 0.049 & \textbf{1.10} & 2.97 & 78.76 & \textbf{0.59} & 2.14 & 23.59 \\
 & 200 & \textbf{0.005} & 0.016 & 0.060 & \textbf{0.002} & 0.010 & 0.018 & \textbf{0.43} & 1.36 & 51.28 & \textbf{0.18} & 0.88 & 14.77 \\ \hline
\multirow{3}{*}{\shortstack[l]{Mean change \\ unknown \\variance}} & 50 & \textbf{0.032} & 0.071 & 0.204 & \textbf{0.025} & 0.087 & 0.088 & \textbf{2.29} & 6.21 & 108.85 & \textbf{1.67} & 4.87 & 45.27 \\
 & 100 & \textbf{0.014} & 0.040 & 0.121 & \textbf{0.008} & 0.037 & 0.044 & \textbf{1.08} & 3.06 & 82.43 & \textbf{0.62} & 2.45 & 31.65 \\
 & 200 & \textbf{0.004} & 0.015 & 0.055 & \textbf{0.002} & 0.009 & 0.017 & \textbf{0.44} & 1.44 & 55.55 & \textbf{0.20} & 0.95 & 19.64 \\ \hline
\multirow{3}{*}{\shortstack[l]{Continuous slope \\change  known \\variance}} & 50 & \textbf{0.001} & 0.001 & 0.102 & \textbf{0.007} & 0.007 & 0.191 & \textbf{1.13} & 1.47 & 431.77 & \textbf{4.36} & 5.01 & 364.85 \\
 & 100 & \textbf{0.000} & 0.001 & 0.096 & \textbf{0.002} & 0.005 & 0.169 & \textbf{0.48} & 1.38 & 379.88 & \textbf{1.76} & 3.69 & 305.57 \\
 & 200 & \textbf{0.000} & 0.001 & 0.083 & \textbf{0.001} & 0.004 & 0.126 & \textbf{0.20} & 1.42 & 300.92 & \textbf{0.67} & 3.47 & 225.18 \\ \hline
\multirow{3}{*}{\shortstack[l]{Discontinuous \\slope change \\known variance}} & 50 & \textbf{0.025} & 0.054 & 0.149 & \textbf{0.036} & 0.096 & 0.104 & \textbf{2.91} & 7.59 & 106.44 & \textbf{2.72} & 6.60 & 33.86 \\
 & 100 & \textbf{0.012} & 0.031 & 0.092 & \textbf{0.013} & 0.047 & 0.065 & \textbf{1.52} & 3.59 & 82.21 & \textbf{1.19} & 3.70 & 24.58 \\
 & 200 & \textbf{0.004} & 0.013 & 0.048 & \textbf{0.003} & 0.014 & 0.036 & \textbf{0.62} & 1.79 & 54.56 & \textbf{0.40} & 1.77 & 17.31 \\ \hline
\multirow{3}{*}{\shortstack[l]{Seasonal change \\ known variance}} & 50 & \textbf{0.039} & 0.071 & 0.262 & \textbf{0.020} & 0.042 & 0.064 & \textbf{2.57} & 5.82 & 74.20 & \textbf{0.74} & 1.67 & 9.37 \\
 & 100 & \textbf{0.017} & 0.038 & 0.152 & \textbf{0.005} & 0.014 & 0.028 & \textbf{1.06} & 2.71 & 56.32 & \textbf{0.25} & 0.63 & 5.48 \\
 & 200 & \textbf{0.005} & 0.013 & 0.066 & \textbf{0.001} & 0.004 & 0.010 & \textbf{0.40} & 1.25 & 36.12 & \textbf{0.08} & 0.27 & 3.19 \\ \hline
\end{tabular}
}
\caption{Simulation on a fixed horizon $T=1000$ with $500$ replicates. $W_{1,\text{change}}$ is the averaged Wasserstein distance conditional on change models, and $\text{TV}_{\text{full}}$ is the averaged total variation distance on the full posterior. The lowest value in each row is bold based on the unrounded values.}
\label{tab:calibration}
\end{table}
Simulation results are shown in Table \ref{tab:calibration}. The approximation accuracy improves as the budget $M$ increases for all three approaches. LP-P shows poor posterior approximation as it has the largest total variation and Wasserstein distance, especially in the continuous slope case. Merging has more accurate posterior approximation compared to pruning, as LP-M and Bagel show smaller error. Moreover, Bagel closely approximates the exact posterior across all examples and has smaller error compared to LP-M. So the merge criterion does help improve the posterior approximation.

\subsection{Accuracy under online setting}\label{compare_benchmark}

We now compare the performance of methods in terms of whether and how quickly they detect a change, and, once they have detected one, how accurate inference about where that change is. We simulate data with a fixed size of change and changepoint location of 1000. We set the pre-change prior mean to match the data-generating process. As before, we consider a main prior and a diffuse one. In each case, the prior for the change size $\gamma$ has mean zero. For the first prior, we choose the post-change variance to be the squared size of the simulated change for the main prior. The diffuse prior uses a covariance matrix ten times this. The specific data generating process and priors are given in \ifarxiv Appendix \ref{Supplement A:sim_priors} \fi. Different from the previous study, each method stops at the first time its detection statistic exceeds a predefined threshold. The threshold is calibrated on data of length 2000 without a change and set to achieve $5\%$ false-alarm rate over 500 replications. 

We look at a range of properties of each method, including the accuracy of the posterior distribution of the location of a change when a change is detected. To do this, for each approximate method, we rerun the exact method on the same data at the stopping time of the approximate method, to obtain the exact posterior. For replicate $r$, denote $T_{a}^r$ as the stopping time, $\hat\tau_{a}^r$ as the MAP estimator, and $H_a^r$ as the 95\% HPD set of the approximate methods. $\hat\tau_e^r(T_a^r)$ and $H_e^r(T_a^r)$ are the MAP estimator and 95\% HPD set of the exact approach evaluated at $T_{a}^r$. We evaluate metrics as in Table \ref{tab:stopping_metrics} shows. The invalid percentage shows the proportion of replicates that trigger false alarms or fail to alarm. The average detection delay provides a reference detection speed. The MAP distance measures the distance between the approximate and exact MAP estimates. The Jaccard index and HPD mass error measure the accuracy of HPD sets at the nominated level.

\begin{table}
    \centering
    \small
    \scalebox{1}{
    \begin{tabular}{l|l}
    \hline
    Metric & Formula \\\hline
    Invalid proportion (Inv) & $1-\frac{|V|}{R}$ \\
    Average detection delay (Delay) & $\frac{1}{|V|}\sum_{r\in V}(T_a^r-\tau)$  \\
    Median MAP distance (MAP dist) & $\operatorname{median}_{r\in V}|\hat{\tau}_a^r(T_a^r)-\hat{\tau}_e^r(T_a^r)|$  \\
    Mean Jaccard index (Jaccard) & $\frac{1}{|V|}\sum_{r\in V}\frac{|H_a^r(T_a^r)\cap H_e^r(T_a^r)|}{|H_a^r(T_a^r)\cup H_e^r(T_a^r)|}$  \\
    HPD mass error (HPD err) & $\frac{100}{|V|}\sum_{r\in V}|P_e(H_a^r(T_a^r)\mid y_{1:T_a^r})-0.95|$ \\
    \hline
    \end{tabular}}
    \caption{Evaluation metrics used in Table \ref{tab:example2}. $V=\{r:\tau< T_a^r\leq 2000\}$ is the valid replicates, $P_e$ is the exact posterior mass and $R$ is the total number of replicates.}
    \label{tab:stopping_metrics}
\end{table}

Table \ref{tab:example2} presents the simulation results for two representative examples under the main prior. Additional results under both main and diffuse priors are provided in \ifarxiv
Appendix \ref{Supplement A:sim_priors}
\fi. Bagel and LP-M have fewer invalid replicates and shorter delays compared to the pruned approach. When evaluating the posterior approximation accuracy, Bagel consistently provides a good approximation, with the lowest HPD error and the highest Jaccard. This indicates that the HPD set inferred by Bagel accurately preserves the changepoint candidates and their corresponding probabilities. On the other hand, LP-P can preserve the accuracy of the MAP estimator in most cases but has poor estimated HPD sets. The lower Jaccard overlap and larger HPD error indicate that LP-P can lead to misleading uncertainty quantification. The performance of LP-M sits between LP-P and Bagel in most cases. It can provide a better approximation than LP-P but has lower Jaccard and larger HPD error than Bagel, and can introduce small shifts in the MAP estimator. In the continuous slope change, Bagel and LP-M have similar performance. This is because neighbouring candidates may already have highly similar post-change distributions due to continuity constraints.

\begin{table}
\centering
\small
\scalebox{0.9}{
\begin{tabular}{l|r|c|c|c|c|c|c}
\hline
 & && \multicolumn{5}{c}{Main prior}\\
\cline{4-8}
Example & $M$ & Method & Inv & Delay & MAP dist & Jaccard & HPD err \\
\hline
\multirow{10}{*}{\shortstack[l]{Mean\\ change \\ unknown \\variance}} & -- & Exact & 0.04 & 282& 0 & 1.00 & 0.02 \\
\cline{2-8}
 & \multirow{3}{*}{50} & Bagel & 0.04 & 282& \textbf{0} & \textbf{0.90} & \textbf{0.27} \\
 &  & LP-M & 0.05 & 279 & 3 & 0.79 & 1.83 \\
 &  & LP-P & 0.08 & 306 & \textbf{0} & 0.14 & 56.68 \\
\cline{2-8}
 & \multirow{3}{*}{100} & Bagel & 0.04 & 283& \textbf{0} & \textbf{0.94} & \textbf{0.10} \\
 &  & LP-M & 0.04 & 282 & 1 & 0.85 & 0.87 \\
 &  & LP-P & 0.06 & 294 & \textbf{0} & 0.25 & 39.10 \\
\cline{2-8}
 & \multirow{3}{*}{200} & Bagel & 0.04 & 282& \textbf{0} & \textbf{0.96} & \textbf{0.05} \\
 &  & LP-M & 0.04 & 283& \textbf{0} & 0.90 & 0.38 \\
 &  & LP-P & 0.05 & 287 & \textbf{0} & 0.42 & 22.01 \\
\hline
\multirow{10}{*}{\shortstack[l]{Discontinuous \\ slope\\ change \\ known \\ variance}} & -- & Exact & 0.02 & 346 & 0 & 1.00 & 0.02 \\
\cline{2-8}
 & \multirow{3}{*}{50} & Bagel & 0.02 & 348 & \textbf{0} & \textbf{0.92} & \textbf{0.48} \\
 &  & LP-M & 0.02 & 351 & \textbf{0} & 0.87 & 1.62 \\
 &  & LP-P & 0.09 & 409 & \textbf{0} & 0.04 & 83.02 \\
\cline{2-8}
 & \multirow{3}{*}{100} & Bagel & 0.02 & 348 & \textbf{0} & \textbf{0.95} & \textbf{0.14} \\
 &  & LP-M & 0.02 & 348 & \textbf{0} & 0.91 & 0.69 \\
 &  & LP-P & 0.05 & 393 & \textbf{0} & 0.09 & 73.89 \\
\cline{2-8}
 & \multirow{3}{*}{200} & Bagel & 0.02 & 346 & \textbf{0} & \textbf{0.97} & \textbf{0.06} \\
 &  & LP-M & 0.02 & 348 & \textbf{0} & 0.94 & 0.30 \\
 &  & LP-P & 0.03 & 383 & \textbf{0} & 0.19 & 59.68 \\\hline
\end{tabular}
}
\caption{Simulation result on a fixed horizon $T=2000$ with changepoint location $\tau=1000$ over 500 replicates. The lowest MAP distance, largest Jaccard index and lowest HPD error are in bold based on unrounded values. }
\label{tab:example2}
\end{table}

\subsection{Empirical runtime}
\begin{figure}
    \centering
    \begin{tikzpicture}
        \node[inner sep=0, anchor=south west] (img)
        at (0,0)
        {\includegraphics[width=0.6\linewidth]{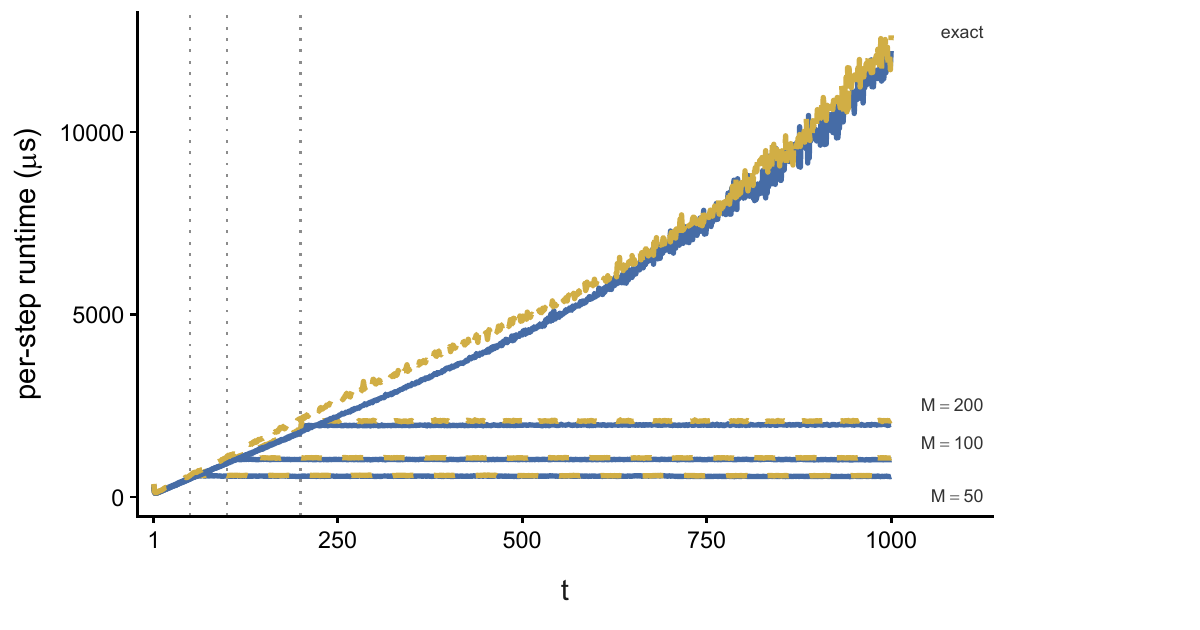}};
        \begin{scope}[shift={(img.south east)}, xshift=-1.4cm, yshift=1cm]
            \draw[
                line width=1.2pt,
                color={rgb,1:red,0.275;green,0.424;blue,0.651}
            ] (0,0.35) -- (0.6,0.35);
            \node[anchor=west, font=\scriptsize] at (0.75,0.35) {Known variance};

            \draw[
                line width=1.2pt,
                color={rgb,1:red,0.820;green,0.682;blue,0.271}
            ] (0,0) -- (0.6,0);
            \node[anchor=west, font=\scriptsize] at (0.75,0) {Unknown variance};
        \end{scope}
    \end{tikzpicture}
    \caption{Median per-step runtime of \Bagel{} on $T=1000$ length of mean change example for $M = 50,100,200$, over 200 replications. The vertical dashed lines represent $t=M$, which is the start time of the merging procedure. The fluctuation at the end of the exact approach is due to its memory footprint.}
    \label{fig:speed_main}
\end{figure}
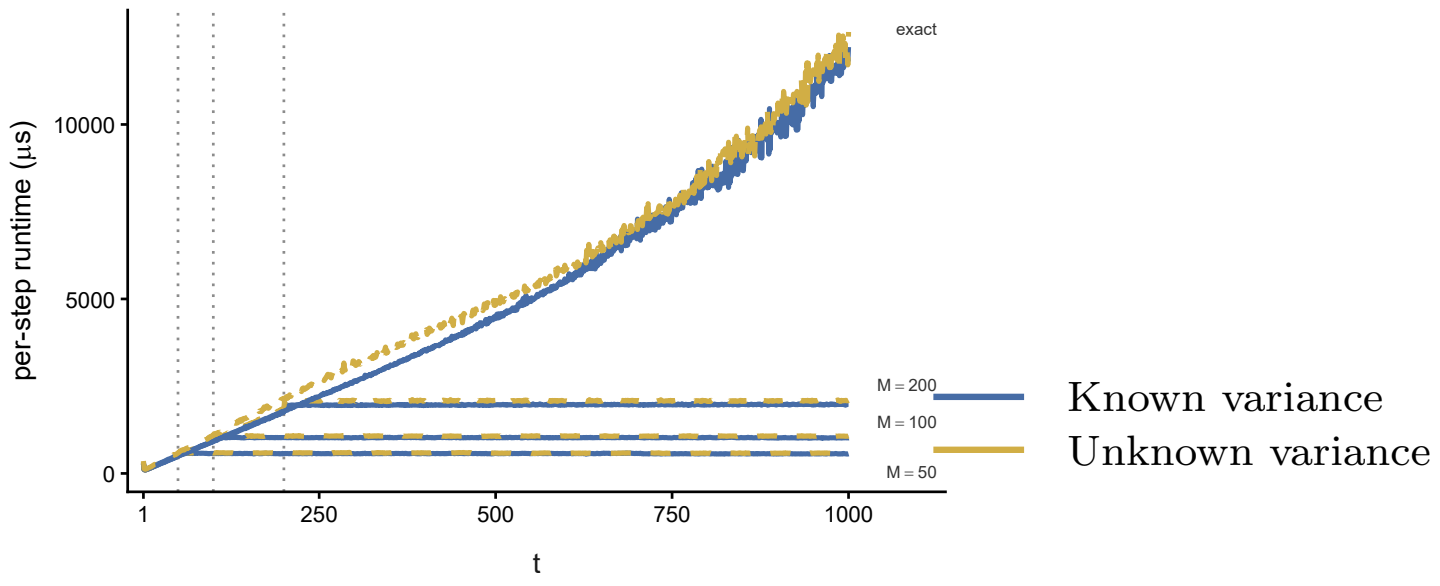

We evaluate the empirical runtime of the exact approach and Bagel. The main code of Bagel is implemented in C++, and invoked per step from the R interface. Figure \ref{fig:speed_main} shows the runtime when varying $M$ in the mean change example. Other settings where $M$ is fixed but the dimension of the design matrix varies are analysed in 
\ifarxiv
Appendix \ref{Supplement A:sim_speed}
\fi. Bagel is substantially faster and more stable than the exact approach over long monitoring horizons. These results show that merging preserves the accuracy of the exact method while remaining fast to implement, making Bagel computationally practical for high-frequency, long-run and real-world monitoring tasks. 

\section{Real data example - Azure LLM inference traces}

Large language model (LLM) inference workloads are dynamic over time. Changes in generated-token count reflect the change decoding demand, which further affects latency. Detecting these changes is useful for workload monitoring, diagnosis and capacity planning. We analyse the generated token count for code traffic from the Azure LLM inference trace \citep{stojkovic2025dynamollm} available from {\texttt{https://github.com/Azure/AzurePublicDataset/}}. The data records the request timestamps and generated-token counts over one week in 2024 from Friday to Thursday. We aggregate the requests into five-minute intervals (2016 bins), and monitor the mean generated-token count for code traffic within each interval as a summary of decoding demand. 
We assume the first day (288 bins) has no change to calibrate the alarming threshold, and the remaining six days are for monitoring. The model is
\begin{eqnarray*}
y_t &=& \beta_0 + \beta_1 \frac{t-1}{288}
+ \beta_{\sin}\sin\frac{2\pi h t}{288}
+ \beta_{\cos}\cos\frac{2\pi h t}{288} \\
& & + \mathbf{1}(t>\tau)\left[
\gamma_0 + \gamma_1\frac{t-\tau}{288}
+ \gamma_{\sin}\sin\frac{2\pi h t}{288}
+ \gamma_{\cos}\cos\frac{2\pi h t}{288}
\right]
+ \sigma\epsilon_t,
\end{eqnarray*}
where $\sigma^2\sim\mathrm{IG}(\nu,\iota)$. We set
$\beta\mid\sigma^2\sim N(0,100\sigma^2 I_4)$ and
$\sigma^2\sim\mathrm{IG}(2,1)$. We calculate the posterior for the pre-change parameters from the training data, and set this to be the prior distribution for monitoring. We further use this posterior to inform our prior for the post-change parameters.
We assume the marginal prior variance of the intercept change and slope change is the posterior noise variance. We also assume, the marginal prior standard deviation of seasonal change is half the estimated seasonal amplitude. 
We sample data with a length of 6 days by drawing the posterior parameters and block-resampled training residuals. The threshold is chosen to achieve 10\% false alarm rates over 1000 replicates. During monitoring, when an alarm is triggered, the changepoint location is fixed at its posterior MAP estimate. Conditional on this location, we recompute the exact posterior parameters for the post-change segment and use this as the prior for pre-change parameters for the next monitoring window whilst keeping the threshold fixed.

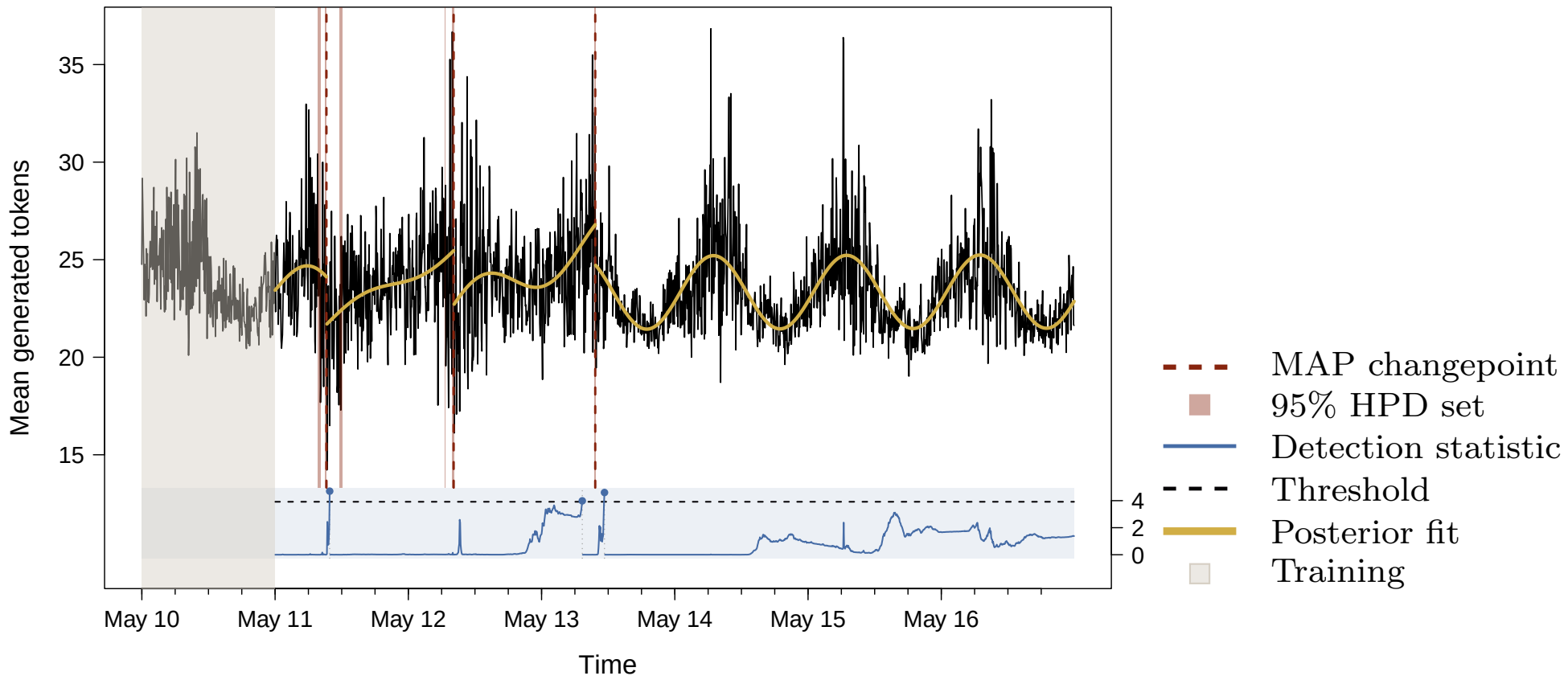
\begin{figure}[]
    \centering
    \begin{tikzpicture}
        \node[inner sep=0, anchor=south west] (img) at (0,0) {
      \includegraphics[width=0.8\linewidth]{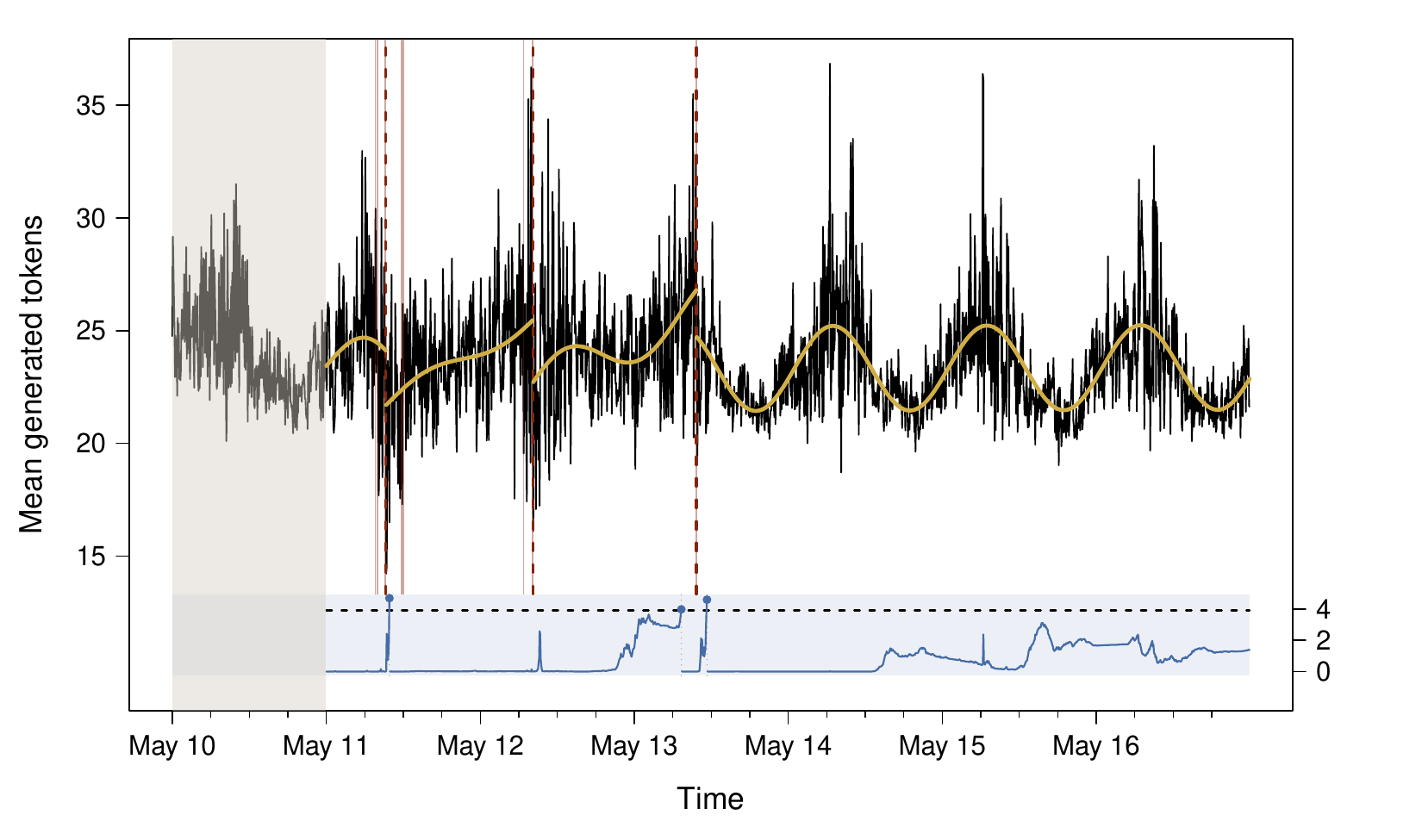}};
      \begin{scope}[shift={(img.south east)}, xshift=-0.5cm, yshift=1.0cm]
            \draw[
                dashed,
                line width=1.2pt,
                color={rgb,1:red,0.529;green,0.141;blue,0.055}
            ] (0,1.75) -- (0.6,1.75);
            \node[anchor=west, font=\scriptsize] at (0.75,1.75) {MAP changepoint};
            \draw[
                draw={rgb,1:red,0.529;green,0.141;blue,0.055},
                fill={rgb,1:red,0.529;green,0.141;blue,0.055},
                fill opacity=0.40,
                draw opacity=0.40
            ] (0.22,1.36) rectangle (0.38,1.52);
            \node[anchor=west, font=\scriptsize] at (0.75,1.44) {95\% HPD set};
            \draw[
                line width=0.9pt,
                color={rgb,1:red,0.275;green,0.424;blue,0.651}
            ] (0,1.10) -- (0.6,1.10);
            \node[anchor=west, font=\scriptsize] at (0.75,1.10) {Detection statistic};
            \draw[
                dashed,
                line width=0.9pt,
                color={rgb,1:red,0.016;green,0.008;blue,0.016}
            ] (0,0.75) -- (0.6,0.75);
            \node[anchor=west, font=\scriptsize] at (0.75,0.75) {Threshold};
            \draw[
                line width=1.8pt,
                color={rgb,1:red,0.820;green,0.682;blue,0.271}
            ] (0,0.40) -- (0.6,0.40);
            \node[anchor=west, font=\scriptsize] at (0.75,0.40) {Posterior fit};
            \draw[
                draw={rgb,1:red,0.839;green,0.812;blue,0.769},
                fill={rgb,1:red,0.839;green,0.812;blue,0.769},
                fill opacity=0.45
            ] (0.22,-0.03) rectangle (0.38,0.13);
            \node[anchor=west, font=\scriptsize] at (0.75,0.05) {Training};
        \end{scope}
    \end{tikzpicture}
    \caption{The detection results of \Bagel on Azure LLM inference trace data with $M=100$. The red dashed line indicates the MAP estimator of changepoint location. The yellow line provides the posterior fit. The blue curve shows the alarm statistic $-\log w_0$, where $w_0$ is the posterior probability of no change. The detector alarms when $-\log w_0 > c$, which is equivalent to $1-w_0 >1 - e^{-c} $.}
    \label{fig:azure}
\end{figure}

Figure \ref{fig:azure} presents the detection results. The MAP estimator of changepoint locations is at bins 400, 675 and 981, with alarms at indices 407, 953 and 1001. The estimated duration between MAP and alarm is 7, 278 and 20 observations. The estimated changepoint locations divide the monitored observations into four segments: early Saturday morning, Saturday morning to Sunday morning, Sunday morning to Monday morning, and Monday through Thursday. The second change is associated with changes in the daily seasonality, so it requires longer periods to accumulate sufficient evidence. The first and third alarms are easier to detect, as both of them have level change as well as seasonality change. Although there is no ground truth label for anomalies, the detected changepoint matches visible shifts in the level and daily pattern.

\section{Discussion}

We propose Bagel, a computationally efficient Bayesian online changepoint detection algorithm for linear models that has the ability to detect changes in mean and slope and can incorporate seasonality. By merging similar neighbouring changepoint candidates, \Bagel limits the number of posterior particles that must be updated so the computational complexity is reduced to a constant per observation after reaching the limit. The merged posterior can be recovered later to provide uncertainty quantification. Through the simulation, we show that Bagel maintains a better approximation of the exact posterior compared to LP-P and LP-M. Therefore, under strict latency requirements and limited computational budgets, \Bagel can provide stable and fast sequential updates while retaining rich posterior information. This is particularly useful in practice, as information about when the change happens and its posterior probability can help downstream root-cause analysis and decision-making, and its efficient computation also makes it practical to deploy without demanding substantial resources. 
\ifarxiv
The Bagel package is available at \url{https://github.com/grosed/bagel.git}.
\fi


\ifarxiv

\paragraph{Acknowledgements}

This work is supported by EPSRC grants EP/N031938/1, EP/Z531327/1, EP/Y028783/1, EP/S022252/1 and BT Group.

\paragraph{AI Declaration:}

During the preparation of this work, Claude Code Opus 4.8 and Codex 5.6-sol were used to provide code assistance for the simulation and real data experiment. Specifically, they were used to assist with code refactoring, drafting, debugging, and formatting the R output to LaTeX for analyses specified by the authors. After using this tool, the first author reviewed and tested the code as needed and takes full responsibility for the final outputs.
\fi

\clearpage
\ifarxiv
\appendix

\section{Properties of the Prior in Section 2.1.}\label{Supplement A:prior}

\noindent
{\bf Example 1}

To see that the second prior for Example 1 gives the same distribution for the pre-change and post-change mean, and that these are independent, we apply a change of variable. We have $(\beta,\gamma)^\top$ has a Gaussian distribution with mean 
$
\left[\begin{array}{c}
       \mu_1 \\
       0 
    \end{array} \right]$ and variance 
  \(
\sigma^2\begin{bmatrix}
    \delta_1 & -\delta_1  \\
    -\delta_1 & 2\delta_1  
\end{bmatrix}.
\)

Let the pre-change and post-change means be $(\beta,\theta)^\top$ then 
\[
\left[\begin{array}{c}
       \beta \\
       \theta
    \end{array} \right] = \begin{bmatrix}
    1 & 0  \\
    1 & 1   
\end{bmatrix}  \left[\begin{array}{c}
       \beta \\
       \gamma
    \end{array} \right].
\]
This is a linear transformation, so $(\beta,\theta)^\top$ will also have a Gaussian distributions with mean
\[
\begin{bmatrix}
    1 & 0  \\
    1 & 1   
\end{bmatrix}  \left[\begin{array}{c}
       \mu_1 \\
       0 
    \end{array} \right] = \left[\begin{array}{c}
       \mu_1 \\
       \mu_1
    \end{array} \right],
\]
and variance
\[ \sigma^2
\begin{bmatrix}
    1 & 0  \\
    1 & 1   
\end{bmatrix} 
\begin{bmatrix}
    \delta_1 & -\delta_1  \\
    -\delta_1 & 2\delta_1  
\end{bmatrix}
\begin{bmatrix}
    1 & 0  \\
    1 & 1   
\end{bmatrix}^\top = \sigma^2 \begin{bmatrix}
    \delta_1 & 0  \\
    0 & \delta_1  
\end{bmatrix},
\]
as required.\\

\noindent
{\bf Example 2}

In the change-in-slope case, first consider the continuous change model. We can re-parameterise the model to the pre-change intercept and slope, the change in the mean at the changepoint and the change in slope. Denote these parameters as $(\beta_1,\beta_2,\psi,\gamma_2)^\top$, it is straightforward to show that they are defined as
\[
\left[\begin{array}{c}
       \beta_1 \\
       \beta_2 \\
       \psi \\
       \gamma_2
    \end{array} \right] = \begin{bmatrix}
    1 & 0 &0 &0 \\
    0 & 1 &0&0\\
    0&0&1&\tau\\
    0&0&0&1
\end{bmatrix}  \left[\begin{array}{c}
       \beta_1 \\
       \beta_2 \\
       \gamma_1\\
       \gamma_2
    \end{array} \right].
\]
We can calculate the mean and variance of $(\beta_1,\beta_2,\psi,\gamma_2)^\top$ as, respectively:
\[
\begin{bmatrix}
    1 & 0 &0 &0 \\
    0 & 1 &0&0\\
    0&0&1&\tau\\
    0&0&0&1
\end{bmatrix} \left[\begin{array}{c}
       \mu_1 \\
       \mu_2\\
       0\\
       0
    \end{array} \right]=\left[\begin{array}{c}
       \mu_1 \\
       \mu_2\\
       0\\
       0
    \end{array} \right], \mbox{ and}
\]
\[
\sigma^2
\begin{bmatrix}
    1 & 0 &0 &0 \\
    0 & 1 &0&0\\
    0&0&1&\tau\\
    0&0&0&1
\end{bmatrix} 
\begin{bmatrix}
\delta_1 & 0 & 0& 0 \\
0 & \delta_2 & 0 & 0 \\
0 & 0& \tau^2\delta_3  &  -\tau\delta_3 \\
0 &  0 & -\tau\delta_3 & \delta_3 
\end{bmatrix}\begin{bmatrix}
    1 & 0 &0 &0 \\
    0 & 1 &0&0\\
    0&0&1&\tau\\
    0&0&0&1 
\end{bmatrix}^\top
=\sigma^2 \begin{bmatrix}
    \delta_1 & 0 &0 &0 \\
    0 & \delta_2 &0&0\\
    0&0&0&0\\
    0&0&0&\delta_3
\end{bmatrix}.
\]
This shows that, $\psi$, the change in mean at the changepoint is equal to zero, and hence the mean function is continuous, and that the change in slope is independent of the pre-change parameters.

Now consider a discontinuous change. This time we re-parameterise in terms of the pre-change and post-change intercept and slope, whih we denote $(\beta_1,\beta_2,\theta_1,\theta_2)^\top$. These satisfy
\[
\left[\begin{array}{c}
       \beta_1 \\
       \beta_2 \\
       \theta_1 \\
       \theta_2
    \end{array} \right] = \begin{bmatrix}
    1 & 0 &0 &0 \\
    0 & 1 &0&0\\
    1&\tau&1&\tau\\
    0&1&0&1
\end{bmatrix}  \left[\begin{array}{c}
       \beta_1 \\
       \beta_2 \\
       \gamma_1\\
       \gamma_2
    \end{array} \right].
\]
With this linear transformation, and using the prior specified in Section 2.1 for the discontinuous model, we have that the mean of $(\beta_1,\beta_2,\theta_1,\theta_2)^T$ is
\[
\begin{bmatrix}
    1 & 0 &0 &0 \\
    0 & 1 &0&0\\
    1&\tau&1&\tau\\
    0&1&0&1
\end{bmatrix} \left[\begin{array}{c}
       \mu_1 \\
       \mu_2\\
       -\mu_2\tau\\
       0
    \end{array} \right]=\left[\begin{array}{c}
       \mu_1 \\
       \mu_2\\
       \mu_1\\
       \mu_2
    \end{array} \right],
\]
and its variance is
\[\sigma^2
\begin{bmatrix}
    1 & 0 &0 &0 \\
    0 & 1 &0&0\\
    1&\tau&1&\tau\\
    0&1&0&1
\end{bmatrix} 
\begin{bmatrix}
\delta_1 & 0 & -\delta_1& 0 \\
0 & \delta_2 & 0 & -\delta_2 \\
-\delta_1 & 0& 2\delta_1+\tau^2\delta_2  &  -\tau\delta_2 \\
0 &  -\delta_2 & -\tau\delta_2 & 2\delta_2 
\end{bmatrix}\begin{bmatrix}
    1 & 0 &0 &0 \\
    0 & 1 &0&0\\
    1&\tau&1&\tau\\
    0&1&0&1
\end{bmatrix}^\top
=\sigma^2 \begin{bmatrix}
    \delta_1 & 0 &0 &0 \\
    0 & \delta_2 &0&0\\
    0&0&\delta_1&0\\
    0&0&0&\delta_2
\end{bmatrix}.\]
So the pre-change and post-change parameters are independent and follow the same distribution, as required.

\noindent
{\bf Example 3}

We consider the prior for Example 3 that the pre-change and post-change distributions are independent and identically distributed. Our prior is that $(\beta_{\sin}, \beta_{\cos},\gamma_{\sin},\gamma_{\cos})^\top$ has a Gaussian distribution with mean $\begin{bmatrix}
    \mu_{\sin}&\mu_{\cos}&0&0
\end{bmatrix}^T$ and variance 
  \[
\sigma^2\begin{bmatrix}
      \delta_{\sin}& 0 &  -\delta_{\sin}&0  \\
        0 & \delta_{\cos}& 0  & -\delta_{\cos}  \\
      -\delta_{\sin}&0  &2\delta_{\sin}& 0 \\
     0&-\delta_{\cos}& 0&2\delta_{\cos}  \\
\end{bmatrix}.
\]
Let the pre-change and post-change coefficients of the $\sin$ and $\cos$ terms be $(\beta_{\sin},\beta_{\cos},\theta_{\sin},\theta_{\cos})^\top$, then 
\[
\left[\begin{array}{c}
       \beta_{\sin} \\
       \beta_{\cos} \\
       \theta_{\sin} \\
       \theta_{\cos} \\
    \end{array} \right] = \begin{bmatrix}
    1 & 0 & 0 &0  \\
    0 & 1 & 0 &0 \\  
    1 & 0 & 1 &0  \\
    0 & 1 & 0 &1   
\end{bmatrix}  \left[\begin{array}{c}
       \beta_{\sin} \\
       \beta_{\cos} \\
       \gamma_{\sin} \\
       \gamma_{\cos} \\
    \end{array} \right].
\]
So $(\beta_{\sin},\beta_{\cos},\theta_{\sin},\theta_{\cos})^\top$ will also have a Gaussian distributions with mean
\[
\begin{bmatrix}
    1 & 0 & 0 &0  \\
    0 & 1 & 0 &0 \\  
    1 & 0 & 1 &0  \\
    0 & 1 & 0 &1   
\end{bmatrix}  \left[\begin{array}{c}
       \mu_{\sin} \\
       \mu_{\cos} \\
       0 \\0 
    \end{array} \right] = \left[\begin{array}{c}
       \mu_{\sin} \\
       \mu_{\cos} \\
       \mu_{\sin} \\
       \mu_{\cos} 
    \end{array} \right],
\]
and variance
\[ \sigma^2
\begin{bmatrix}
    1 & 0 & 0 &0  \\
    0 & 1 & 0 &0 \\  
    1 & 0 & 1 &0  \\
    0 & 1 & 0 &1   
\end{bmatrix}\begin{bmatrix}
      \delta_{\sin}& 0 &  -\delta_{\sin}&0  \\
        0 & \delta_{\cos}& 0  & -\delta_{\cos}  \\
      -\delta_{\sin}&0  &2\delta_{\sin}& 0 \\
     0&-\delta_{\cos}& 0&2\delta_{\cos}  \\
\end{bmatrix}
\begin{bmatrix}
    1 & 0 & 0 &0  \\
    0 & 1 & 0 &0 \\  
    1 & 0 & 1 &0  \\
    0 & 1 & 0 &1   
\end{bmatrix}^\top = \sigma^2\begin{bmatrix}
      \delta_{\sin}& 0 &  0&0  \\
        0 & \delta_{\cos}& 0  & 0  \\
      0&0  &\delta_{\sin}& 0 \\
     0&0& 0&\delta_{\cos}  \\
\end{bmatrix},
\]
as required.\\
\section{Proofs for sequential updating}\label{sequential_update}

{\bf Proof of Theorem 2.1}\\
For $i=1,\ldots,t-2$ the recursion comes from the definition of the weights.
\begin{eqnarray*}
w_{i,t}&=&\Pr_t(\tau=i)p(y_{1:t}|\tau=i) \\&=&\Pr_{t-1}(\tau=i) p(y_{1:t-1}|\tau=i) \frac{\Pr_t(\tau=i)}{\Pr_{t-1}(\tau=i)}p(y_t|y_{1:t-1},\tau=i)\\
&=& w_{i,t-1} \frac{\Pr_t(\tau=i)}{\Pr_{t-1}(\tau=i)}p(y_t|y_{1:t-1},\tau=i).
\end{eqnarray*}
A similar simple argument applies for $i=0$. Finally for $i=t-1$ we have
\begin{eqnarray*}
w_{t-1,t}&=&\Pr_t(\tau=t-1)p(y_{1:t}|\tau=t-1)\\&=&\Pr_{t-1}(\tau\geq t-1) p(y_{1:t-1}|\tau\geq t-1) \frac{\Pr_t(\tau=i)}{\Pr_{t-1}(\tau\geq t-1)}p(y_t|y_{1:t-1},\tau=t-1)\\
&=& w_{0,t-1} \frac{\Pr_t(\tau=i)}{\Pr_{t-1}(\tau\geq t-1)})p(y_t|y_{1:t-1},\tau=t-1).
\end{eqnarray*}

\noindent
{\bf Proof of Theorem 2.2}\\
If we have a change at $t-1$ then the prior for $(\beta,\gamma)$ has mean $\mu^{t-1}$ and variance $\sigma^2\Sigma^{t-1}$. Our assumption is that the marginal prior for $\beta$ is the same as that assuming no change, but the conditional prior for $\gamma$ given $\beta$ can depend on the change location $t-1$.

The posterior distribution given $y_{1:t-1}$, conditional on a change at $t-1$ is 
\begin{eqnarray*}
\lefteqn{p(\beta,\gamma,\sigma|y_{1:t-1},\tau=t-1) }\\
&\propto&
p(\sigma)p(\beta|\sigma)p(\gamma|\beta,\sigma,\tau=t-1) p(y_{1:t-1}|\beta,\gamma,\sigma,\tau=t-1)\\
&=& p(\sigma)p(\beta|\sigma)p(\gamma|\beta,\sigma,\tau=t-1) p(y_{1:t-1}|\beta,\sigma,\tau=t-1)\\
&\propto& p(\beta,\sigma|y_{1:t-1},\tau=t-1) p(\gamma|\beta,\sigma,\tau=t-1)  \\
&=& p(\beta,\sigma|y_{1:t-1},\tau\geq t-1) p(\gamma|\beta,\sigma,\tau= t-1).
\end{eqnarray*}
Here we have used the fact that the observations prior to the change do not depend on $\gamma$, and that as the prior for $\beta$ and $\sigma$ and the likelihood for $y_{1:t-1}$ are the same for $\tau=t-1$ and $\tau\geq t-1$ so are the posteriors.

The posterior for $(\beta,\sigma)$ given $y_{1:t-1}$ and $\tau\geq t-1$ is calculated at time $t-1$ for the case of no change by $t-1$ and conditional on $\sigma$ the posterior for $\beta$ is Gaussian with mean $\mu^{t-1,0}$ 
and variance $\Sigma^{t-1,0}$.

As $p(\beta,\gamma|\sigma,\tau=t-1)$ is multivariate Gaussian, the conditional prior $p(\gamma|\beta,\sigma,\tau= t-1)$ is also Gaussian, with a mean that is linear in $\beta$. Using standard results for the multivariate Gaussian gives that this conditional distribution is of the form
\[
\gamma|\beta,\sigma, \tau=t-1 \sim \mbox{N}( \mu_{\gamma}+\Sigma_{\gamma,\beta}\Sigma_{\beta,\beta}^{-1}(\beta-\mu_\beta), \sigma^2 (\Sigma_{\gamma,\gamma}-\Sigma_{\gamma,\beta}\Sigma_{\beta,\beta}^{-1}\Sigma_{\beta,\gamma})),
\]

Therefore, the marginal mean of $\gamma$ after updating $\beta$ is given by
\begin{align*}
&E(\gamma) = E\left(E(\gamma \mid \beta)\right) = \mu_\gamma + \Sigma_{\gamma,\beta}\Sigma_{\beta,\beta}^{-1}(\mu^{t-1,0} - \mu_\beta).
\end{align*}
The marginal scaled variance is
\begin{align*}
\mbox{Var}(\gamma) &= E\left(\mbox{Var}(\gamma \mid \beta)\right) + \mbox{Var}\left(E(\gamma \mid \beta)\right)\\
&=\sigma^2\left(\Sigma_{\gamma,\gamma}-\Sigma_{\gamma,\beta}\Sigma_{\beta,\beta}^{-1}\Sigma_{\beta,\gamma}+\Sigma_{\gamma,\beta}\Sigma_{\beta,\beta}^{-1}\Sigma^{t-1,0}\Sigma_{\beta,\beta}^{-1}\Sigma_{\beta,\gamma} \right)\\
&=\sigma^2\left(\Sigma_{\gamma,\gamma}+\Sigma_{\gamma,\beta}\Sigma_{\beta,\beta}^{-1}(\Sigma^{t-1,0}-\Sigma_{\beta,\beta})\Sigma_{\beta,\beta}^{-1}\Sigma_{\beta,\gamma}\right).
\end{align*}
The covariance can be calculated as
\[
\mbox{Cov}( \gamma,\beta) = \sigma^2\mbox{Cov}\left(\Sigma_{\gamma,\beta}\Sigma_{\beta,\beta}^{-1}\beta,  \beta\right) = \sigma^2\Sigma_{\gamma,\beta}\Sigma_{\beta,\beta}^{-1} \Sigma^{t-1,0} .
\]

Thus, the full updated joint distribution of $(\beta, \gamma)$ is
\[
\begin{aligned}
N\Bigg(
&\begin{bmatrix}
\mu^{t-1,0} \\
\mu_\gamma + \Sigma_{\gamma,\beta} \Sigma_{\beta,\beta}^{-1}
(\mu^{t-1,0} - \mu_\beta)
\end{bmatrix},\\
&\sigma^2
\begin{bmatrix}
\Sigma^{t-1,0} &
\Sigma^{t-1,0} \Sigma_{\beta,\beta}^{-1} \Sigma_{\beta,\gamma} \\
\Sigma_{\gamma,\beta} \Sigma_{\beta,\beta}^{-1} \Sigma^{t-1,0} &
\Sigma_{\gamma,\gamma} + \Sigma_{\gamma,\beta}
\Sigma_{\beta,\beta}^{-1}
(\Sigma^{t-1,0} - \Sigma_{\beta,\beta})
\Sigma_{\beta,\beta}^{-1} \Sigma_{\beta,\gamma}
\end{bmatrix}
\Bigg).
\end{aligned}
\]
as required.

\noindent
{\bf Proof of Theorems 2.3 and 2.4}\\
The results in Theorems 2.3 and 2.4 follow from standard results. One way to see this is to view our model as a dynamic linear model \citep{west2006bayesian}, where the hidden state is the parameter, but we define the dynamics such that the hidden state does not change over time.

For the case of no change, the dynamic linear model is
\begin{eqnarray*}
\beta_t&=&\beta_{t-1} \\
y_t &=& h_0^T\beta_t + \sigma \epsilon_t,
\end{eqnarray*}
where $\epsilon_t \sim \mathcal{N}(0,1)$, and $h_0$ is the known regression vector. At time $t-1$, given observations $y_{1:t-1}$, the posterior distribution is:
\[
\beta_{t-1} \mid \sigma^2, y_{1:t-1}, \tau \geq t-1 \sim \mathcal{N}(\mu^{t-1,0}, \sigma^2 \Sigma^{t-1,0}),
\]
\[
\sigma^2 \mid y_{1:t-1}, \tau \geq t-1 \sim IG(\nu^{t-1,0}, \iota^{t-1,0}).
\]
After observing $y_t$, the posterior distribution of $(\beta, \sigma^2)$ is updated using the following formulas as show in \citep{west2006bayesian, fearnhead2011efficient}:
\begin{align*}
\Sigma^{t,0} &= \Sigma^{t-1,0} - A A^\top Q, \\ 
\mu^{t,0} &= \mu^{t-1,0} + A e_0, \\
\nu^{t,0} &= \nu^{t-1,0} + \frac{1}{2}, \\
\iota^{t,0} &= \iota^{t-1,0} + \frac{e_0^2}{2Q},
\end{align*}
where $e_0 = y_t - h_0^\top \mu^{t-1,0}$ is the one-step-ahead forecast error, $Q = h_0^\top \Sigma^{t-1,0} h_0 + 1$ is the forecast error variance and $A = \Sigma^{t-1,0} h_0 / Q$ is the adaptive gain. Then the one-step-ahead predictive distribution of $y_t$ given $y_{1:t-1}$ is:
\[
y_t \mid y_{1:t-1}, \tau \geq t-1 \sim t_{2\nu^{t-1,0}} \left( h_0^\top \mu^{t-1,0}, \frac{\iota^{t-1,0}}{\nu^{t-1,0}} \left( h_0^\top \Sigma^{t-1,0} h_0 + 1 \right) \right),
\]
where $t_{2\nu^{t-1,0}}$ denotes the Student-$t$ distribution with $2\nu^{t-1,0}$ degrees of freedom.

A similar derivation works for the case where there is a change at time $i$, but now the dynamic linear model is
\begin{eqnarray*}
(\beta_t^\top,\gamma_t^\top)^\top&=&(\beta_{t-1}^\top,\gamma_{t-1}^\top)^\top \\
y_t &=& h_i^T(\beta_t^\top,\gamma_t^\top)^\top + \sigma \epsilon_t.
\end{eqnarray*}
The model follows the same dynamic linear structures, but with an augmented parameter vector. As such, the posterior and predictive quantities follow immediately by applying the standard DLM update formulas introduced earlier.

\section{Proof of Proposition 1}\label{Supplement A:gaussian}
For candidate $i$ and its neighbour $j=i+1$ with variance $\sigma^2\Sigma_i <\sigma^2\Sigma_j$, twice the total variation between the two posterior distributions is the coloured region in Figure \ref{fig:gaussian_intersections}.

\begin{figure}
\centering
\begin{tikzpicture}
\begin{axis}[
width=0.65\linewidth,height=5cm,xtick=\empty,ytick=\empty,domain=-4:4,samples=200,ymin=0,ymax=0.45,clip=false,axis lines=none]
\addplot[thick,blue,name path=fj]{1/sqrt(2*pi*4) * exp(-x^2/(2*4))}node[pos=0.82, above] {$f_j(\theta)$};
\addplot[thick,red,name path=fi]{1/sqrt(2*pi) * exp(-x^2/2)}node[pos=0.62, above right] {$f_i(\theta)$};
\addplot[gray,opacity=0.3,forget plot] fill between[of=fi and fj];
\addplot[dashed] coordinates {(-1.36,0) (-1.36,0.159)};
\addplot[dashed] coordinates {( 1.36,0) ( 1.36,0.159)};
\node[below] at (axis cs:-1.36,0) {$c_1$};
\node[below] at (axis cs: 1.36,0) {$c_2$};
\end{axis}
\end{tikzpicture}
\caption{Illustration of two univariate Gaussian densities with
$\sigma^2\Sigma_i<\sigma^2\Sigma_j$.}
\label{fig:gaussian_intersections}
\end{figure}
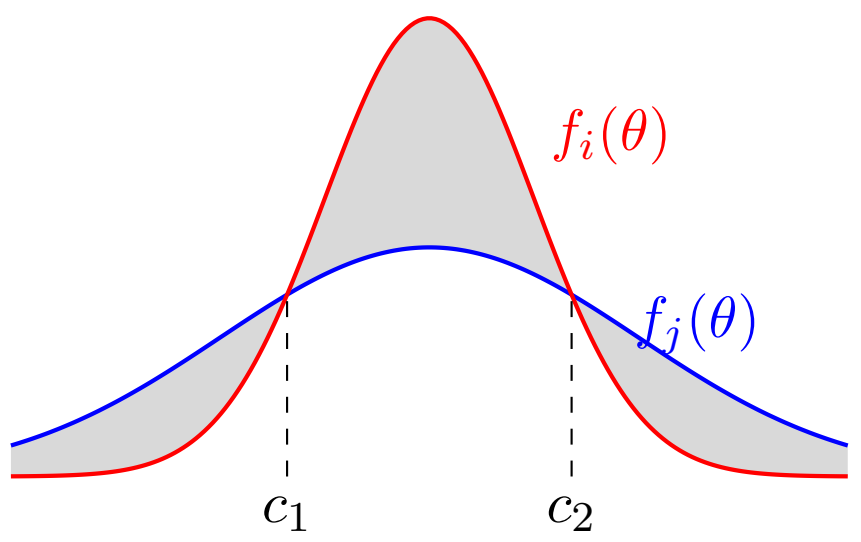
\begin{lemma}
Denote $f_i(\theta)$ and $f_j(\theta)$ as the density of $N(\mu_i,\sigma^2\Sigma_i)$ and $N(\mu_{j},\sigma^2\Sigma_j)$. $f_i(\theta)$ and $f_j(\theta)$ intersect at
\begin{align*}
    (c_1, c_2) = \frac{\mu_i \sigma^2\Sigma_j - \mu_j \sigma^2\Sigma_i \mp \sqrt{\sigma^2\Sigma_i\sigma^2\Sigma_j}\sqrt{(\mu_i - \mu_j)^2 +(\sigma^2\Sigma_j - \sigma^2\Sigma_i)\log \frac{\sigma^2\Sigma_j}{\sigma^2\Sigma_i} }}{\sigma^2\Sigma_j - \sigma^2\Sigma_i}.
\end{align*}
\begin{proof}
The intersection points satisfy $f_i(\theta)=f_j(\theta)$. So,
\begin{align*}
\frac{1}{\sqrt{\sigma^2\Sigma_i}}
\exp\left(
-\frac{(\theta-\mu_i)^2}{2\sigma^2\Sigma_i}
\right)=
\frac{1}{\sqrt{\sigma^2\Sigma_j}}
\exp\left(
-\frac{(\theta-\mu_j)^2}{2\sigma^2\Sigma_j}
\right).
\end{align*}
Taking logarithms we have
\begin{align*}
\frac{(\theta-\mu_j)^2}{\sigma^2\Sigma_j}
-\frac{(\theta-\mu_i)^2}{\sigma^2\Sigma_i}+
\log\frac{\sigma^2\Sigma_j}{\sigma^2\Sigma_i}=0.
\end{align*}
Solving this quadratic equation in $\theta$ with the quadratic formula provides the two roots $c_1$ and $c_2$ stated above.
\end{proof}
\end{lemma}
Therefore, the total variation distance can be written as
\begin{align*}
&TV(f_i(\theta),f_j(\theta))\\
=&\frac{1}{2} \int_{-\infty}^{\infty}|f_i(\theta)-f_j(\theta)|d\theta\\
=&\frac{1}{2}\left[
F_j(c_1)-F_i(c_1)+F_i(c_2)-F_i(c_1)-F_j(c_2)+F_j(c_1)+F_i(c_2)-F_j(c_2)\right] \\
=&F_j(c_1)-F_j(c_2)+F_i(c_2)-F_i(c_1),
\end{align*}
where $F_i(\theta)=\Phi\left(\frac{\theta-\mu_i}{\sqrt{\sigma^2\Sigma_i}}\right)$ and $
F_j(\theta)=\Phi\left(\frac{\theta-\mu_j}{\sqrt{\sigma^2\Sigma_j}}\right)$ as corresponding Gaussian distribution functions. Thus, the total variation between two normal distributions is
\begin{align}\label{TV_uni_gaussian}
     \left[\Phi (\frac{c_1 - \mu_j}{\sqrt{\sigma^2\Sigma_j}}) - \Phi (\frac{c_2 - \mu_j}{\sqrt{\sigma^2\Sigma_j}}) + \Phi (\frac{c_2 - \mu_i}{\sqrt{\sigma^2\Sigma_i}}) - \Phi (\frac{c_1 - \mu_i}{\sqrt{\sigma^2\Sigma_i}})\right].
\end{align}

\section{Priors for the Section 3.1}\label{app:calibration_detail}
For the calibration experiment, the changepoint location follows a uniform distribution $\tau \sim \mathrm{U}(200,800)$ to avoid boundary effects. Within each scenario, we draw the model parameters and noise variances from the same priors as the generated data distribution, so the exact posterior is correctly specified by design. We set the prior probability of no change $\Pr(\tau \ge 1000) = 0.9$. The specific prior setting for each example is as follows.

For mean change with known variance case, data follow $y_t \sim N(\beta, 1)$ for
  $t \le \tau$ and $y_t \sim N(\beta + \gamma, 1)$ for $t > \tau$, where $(\beta, \gamma) \sim N\left(\mathbf{0},\ 0.25^2 \left[\begin{smallmatrix} 1 & -1 \\ -1 & 2\end{smallmatrix}\right]\right)$ for the main setting and $(\beta, \gamma) \sim N\left(\mathbf{0},0.25^2\times10\left[\begin{smallmatrix} 1 & -1 \\ -1 & 2 \end{smallmatrix}\right]\right)$ for the diffuse setting.
  
For mean change with unknown variance case, data follow $y_t \sim N(\beta, \sigma^2)$ for $t \le \tau$ and $y_t \sim N(\beta + \gamma, \sigma^2)$ for $t > \tau$, where $\sigma^2 \sim \mathrm{IG}(6, 5)$ and $(\beta, \gamma) \mid \sigma^2 \sim N\left(\mathbf{0}, 0.25^2\sigma^2 \left[\begin{smallmatrix} 1 & -1 \\ -1 & 2 \end{smallmatrix}\right]\right)$ for the main setting, and $(\beta, \gamma) \mid \sigma^2 \sim N\left(\mathbf{0},\ 10\times 0.25^2\sigma^2 \left[\begin{smallmatrix} 1 & -1 \\ -1 & 2 \end{smallmatrix}\right]\right)$ for the diffuse setting.

For continuous change in slope with known variance case, data follow $y_t = \beta_1 + \beta_2 \Tilde{t} + \varepsilon_t$ for $t \le \tau$, with $\varepsilon_t \sim N(0, 1)$, where $\Tilde{t}$ is the normalised time $\Tilde{t} = t/T$, and $y_t = \beta_1 + \beta_2 \Tilde{t} + \gamma_2 (\Tilde{t} - \Tilde{\tau}) + \varepsilon_t$ for $t > \tau$. The priors are
\[(\beta_1,\beta_2,\gamma_1,\gamma_2)^\top\sim N\left(\mathbf{0},v\left[
\begin{smallmatrix}
1 & 0 & 0 & 0\\
0 & 1 & 0 & 0\\
0 & 0 & \tilde{\tau}^2 & -\tilde{\tau}\\
0 & 0 & -\tilde{\tau} & 1
\end{smallmatrix}\right]\right),
\]
where $v=0.25^2$ for the main setting and $v=10\times 0.25^2$ for the diffuse setting.

For the discontinuous change in slope with known variance case, data follow $y_t = \beta_1 + \beta_2 \Tilde{t} + \varepsilon_t$ for $t \le \tau$, with $\varepsilon_t \sim N(0, 1)$, where $\Tilde{t}$ is the normalised time $\Tilde{t} = t/T$, and $y_t = \ell + s\,(\Tilde{t} - \Tilde{\tau}) + \varepsilon_t$ for $t > \tau$. The priors follow
\[(\beta_1,\beta_2,\gamma_1,\gamma_2)^\top\sim N\left(
\mathbf{0},v\left[\begin{smallmatrix}
1 & 0 & -1 & 0\\
0 & 1 & 0 & -1\\
-1 & 0 & 2+\tilde{\tau}^2 & -\tilde{\tau}\\
0 & -1 & -\tilde{\tau} & 2
\end{smallmatrix}\right]
\right),
\]where $v=0.25^2$ for the main setting and $v=10\times 0.25^2$ for the diffuse setting.

For change in seasonality within the known-variance case, data follow $y_t = \beta_s \sin(2\pi t / 24) + \beta_c \cos(2\pi t / 24) + 
\left[
\gamma_s\sin(2\pi t/24)
+\gamma_c\cos(2\pi t/24)
\right]\mathbf{1}\{t>\tau\}
+\varepsilon_t,$ with $\varepsilon_t \sim N(0, 1)$,The prior is
\[(\beta_s,\beta_c,\gamma_s,\gamma_c)^\top\sim
N\left(\mathbf{0},v\left[
\begin{smallmatrix}
1 & 0 & -1 & 0\\
0 & 1 & 0 & -1\\
-1 & 0 & 2 & 0\\
0 & -1 & 0 & 2
\end{smallmatrix}\right]
\right),
\]
where $v=0.25^2$ for the main setting and $v=10(0.25^2)$ for the diffuse setting.
\section{Priors and full results for the Section 3.2}\label{Supplement A:sim_priors}
\subsection{Prior settings}
In this simulation, we simulate data with a length of 2000 and have a changepoint at 1000. The prior probability of no change is set at $Pr(\tau \geq 2000) = 0.9$.

For the change-in-mean with known variance case, the data follow $N(0,1)$ before change and $N(0.25,1)$ after the change. We set the prior mean of $(\beta,\gamma)$ to be $(0,0)$. The prior variance is $ 0.25^2 \left[\begin{smallmatrix} 1 & -1 \\ -1 & 2\end{smallmatrix}\right]$ for the main setting and set $10 \times 0.25^2 \left[\begin{smallmatrix} 1 & -1 \\ -1 & 2\end{smallmatrix}\right]$ as the diffuse setting.

For the change-in-mean with unknown variance case, the data follow $N(0,1)$ before the change and $N(0.25,1)$ after the change. The prior mean of $(\beta,\gamma)$ is $(0,0)$. The prior variance follows $v \sigma^2 \left[\begin{smallmatrix} 1 & -1 \\ -1 & 2 \end{smallmatrix}\right]$ where $\sigma^2 \sim \mathrm{IG}(6, 5)$. $v=0.25^2$ for the main setting, and $v=10\times0.25^2$ in the diffuse setting.

For the continuous slope change with known variance case, data follow $N(0,1)$ before the change and $1.73(\tilde{t}-0.5)+N(0,1)$ after the change, where $\tilde{t} = t/T$. The prior mean of $(\beta_1,\beta_2,\gamma_1,\gamma_2)$ is $(0,0,0,0)$. The prior variance is $\left[
\begin{smallmatrix}
0.25^2 & 0 & 0 & 0\\
0 & 0.25^2 & 0 & 0\\
0 & 0 & \tilde{\tau}^{2}1.73^2 & -\tilde{\tau}1.73^2\\
0 & 0 & -\tilde{\tau}1.73^2 & 1.73^2
\end{smallmatrix}\right]$ in the main setting and ten times in the diffuse setting.

For the discontinuous slope change with known variance case, data follow $N(0,1)$ before the change and $0.15+1.65(\tilde{t}-0.5)+N(0,1)$, where $\tilde{t} = t/T$. The prior mean of $(\beta_1,\beta_2,\gamma_1,\gamma_2)$ is $(0,0,0,0)$. The prior variance is $\left[\begin{smallmatrix}
0.15^2 & 0 & -0.15^2 & 0\\
0 & 1.65^2 & 0 & -1.65^2\\
-0.15^2 & 0 & 2 \times 0.15^2+1.65^2\tilde{\tau}^{2} & -\tilde{\tau}1.65^2\\
0 & -1.65^2 & -\tilde{\tau}1.65^2 & 2\times1.65^2
\end{smallmatrix}\right]$ in the main setting and ten times in the diffuse setting.

For the seasonality change with known variance case, data follow $y_t=\sin(2\pi t / 24) + \cos(2\pi t / 24) + \left[0.25\sin(2\pi t/24)+0.25\cos(2\pi t/24)\right]\mathbf{1}\{t>\tau\}+\varepsilon_t$ where $\varepsilon_t \sim N(0,1)$. The prior means for $(\beta_s,\beta_c,\gamma_s,\gamma_c)$ is set to be $(1,1,0,0)$, and the prior variance is $v\left[
\begin{smallmatrix}
1 & 0 & -1 & 0\\
0 & 1 & 0 & -1\\
-1 & 0 & 2 & 0\\
0 & -1 & 0 & 2
\end{smallmatrix}\right]$ with $v=0.25^2$ for the main setting and $v=10(0.25^2)$ for the diffuse setting.

\subsection{Results}
Table \ref{appendix:example2} provides the full results for five examples under both main and diffuse priors here. Denote $V=\{r:\tau< T_a^r\leq 2000\}$ as the valid replicates, $P_e$ as the exact posterior mass, and $R$ as the total number of replicates. We compute the invalid proportion ($1-\frac{|V|}{R}$), the average detection delay ($\frac{1}{|V|}\sum_{r\in V}(T_a^r-\tau)$), the median MAP distance ($\operatorname{median}_{r\in V}|\hat{\tau}_a^r(T_a^r)-\hat{\tau}_e^r(T_a^r)|$), the mean Jaccard index ($\frac{1}{|V|}\sum_{r\in V}\frac{|H_a^r(T_a^r)\cap H_e^r(T_a^r)|}{|H_a^r(T_a^r)\cup H_e^r(T_a^r)|}$), and the HPD mass error ($\frac{100}{|V|}\sum_{r\in V}|P_e(H_a^r(T_a^r)\mid y_{1:T_a^r})-0.95|$). 

\begin{table}
\centering
\small
\setlength{\tabcolsep}{4pt}
\renewcommand{\arraystretch}{1} 
\resizebox{\linewidth}{!}{%
\begin{tabular}{l|r|c|c|c|c|c|c|c|c|c|c|c}
\hline
 & & & \multicolumn{5}{c|}{Main prior} & \multicolumn{5}{c}{Diffuse prior} \\
\cline{4-13}
Example & $K$ & Method & Inv. & Delay & MAP dist. & Jaccard & HPD err & Inv. & Delay & MAP dist.  & Jaccard & HPD err \\
\hline
\multirow{10}{*}{\shortstack[l]{Mean \\ change \\ known \\variance}} & -- & Exact & 0.05 & 281 & 0 & 1.00 & 0.02 & 0.09 & 339 & 0 & 1.00 & 0.04 \\
\cline{2-13}
 & \multirow{3}{*}{50} & Bagel & 0.04 & 282 & \textbf{0} & \textbf{0.90} & \textbf{0.26} & 0.09 & 346 & \textbf{0} & \textbf{0.89} & \textbf{0.28} \\
 &  & LP-M & 0.05 & 283 & 3 & 0.79 & 1.85 & 0.10 & 351 & 1 & 0.75 & 3.05 \\
 &  & LP-P & 0.08 & 310 & \textbf{0} & 0.14 & 56.78 & 0.14 & 355 & \textbf{0} & 0.21 & 45.92 \\
\cline{2-13}
 & \multirow{3}{*}{100} & Bagel & 0.05 & 282 & \textbf{0} & \textbf{0.94} & \textbf{0.10} & 0.09 & 341 & \textbf{0} & \textbf{0.94} & \textbf{0.13} \\
 &  & LP-M & 0.05 & 282 & 1 & 0.85 & 0.88 & 0.09 & 346 & \textbf{0} & 0.83 & 1.31 \\
 &  & LP-P & 0.06 & 300 & \textbf{0} & 0.25 & 39.28 & 0.12 & 347 & \textbf{0} & 0.35 & 29.45 \\
\cline{2-13}
 & \multirow{3}{*}{200} & Bagel & 0.05 & 281 & \textbf{0} & \textbf{0.96} & \textbf{0.05} & 0.09 & 339 & \textbf{0} & \textbf{0.97} & \textbf{0.08} \\
 &  & LP-M & 0.04 & 283 & \textbf{0} & 0.90 & 0.36 & 0.09 & 340 & \textbf{0} & 0.90 & 0.52 \\
 &  & LP-P & 0.05 & 294 & \textbf{0} & 0.43 & 21.61 & 0.11 & 349 & \textbf{0} & 0.55 & 15.54 \\
\hline
\multirow{10}{*}{\shortstack[l]{Mean\\ change \\ unknown \\variance}} & -- & Exact & 0.04 & 282 & 0 & 1.00 & 0.02 & 0.09 & 336 & 0 & 1.00 & 0.03 \\
\cline{2-13}
 & \multirow{3}{*}{50} & Bagel & 0.04 & 282 & \textbf{0} & \textbf{0.90} & \textbf{0.27} & 0.09 & 336 & \textbf{0} & \textbf{0.89} & \textbf{0.28} \\
 &  & LP-M & 0.05 & 279 & 3 & 0.79 & 1.83 & 0.09 & 337 & 1 & 0.75 & 3.05 \\
 &  & LP-P & 0.08 & 306 & \textbf{0} & 0.14 & 56.68 & 0.12 & 346 & \textbf{0} & 0.20 & 46.00 \\
\cline{2-13}
 & \multirow{3}{*}{100} & Bagel & 0.04 & 283 & \textbf{0} & \textbf{0.94} & \textbf{0.10} & 0.09 & 334 & \textbf{0} & \textbf{0.94} & \textbf{0.13} \\
 &  & LP-M & 0.04 & 282 & 1 & 0.85 & 0.87 & 0.09 & 339 & \textbf{0} & 0.83 & 1.29 \\
 &  & LP-P & 0.06 & 294 & \textbf{0} & 0.25 & 39.10 & 0.11 & 339 & \textbf{0} & 0.35 & 29.66 \\
\cline{2-13}
 & \multirow{3}{*}{200} & Bagel & 0.04 & 282 & \textbf{0} & \textbf{0.96} & \textbf{0.05} & 0.09 & 336 & \textbf{0} & \textbf{0.97} & \textbf{0.08} \\
 &  & LP-M & 0.04 & 283 & \textbf{0} & 0.90 & 0.38 & 0.09 & 336 & \textbf{0} & 0.90 & 0.53 \\
 &  & LP-P & 0.05 & 287 & \textbf{0} & 0.42 & 22.01 & 0.10 & 335 & \textbf{0} & 0.54 & 15.56 \\
\hline
\multirow{10}{*}{\shortstack[l]{Continuous\\ slope \\ change\\ known \\ variance}} & -- & Exact & 0.02 & 421 & 0 & 1.00 & 0.01 & 0.02 & 505 & 0 & 1.00 & 0.02 \\
\cline{2-13}
 & \multirow{3}{*}{50} & Bagel & 0.02 & 420 & 8 & \textbf{0.95} & 0.30 & 0.02 & 506 & 4 & \textbf{0.95} & \textbf{0.32} \\
 &  & LP-M & 0.02 & 422 & \textbf{6} & 0.95 & \textbf{0.18} & 0.02 & 505 & \textbf{3} & 0.94 & 0.35 \\
 &  & LP-P & 0.04 & 502 & 127 & 0.04 & 92.23 & 0.05 & 559 & 68 & 0.06 & 88.47 \\
\cline{2-13}
 & \multirow{3}{*}{100} & Bagel & 0.02 & 420 & \textbf{3} & \textbf{0.98} & \textbf{0.07} & 0.02 & 505 & \textbf{2} & \textbf{0.98} & \textbf{0.10} \\
 &  & LP-M & 0.02 & 421 & \textbf{3} & 0.97 & 0.12 & 0.02 & 505 & \textbf{2} & 0.97 & 0.23 \\
 &  & LP-P & 0.03 & 451 & 52 & 0.10 & 83.55 & 0.04 & 540 & 6 & 0.14 & 74.11 \\
\cline{2-13}
 & \multirow{3}{*}{200} & Bagel & 0.02 & 420 & 2 & \textbf{0.99} & \textbf{0.03} & 0.02 & 505 & 1 & \textbf{0.99} & \textbf{0.09} \\
 &  & LP-M & 0.02 & 420 & 1 & 0.98 & 0.12 & 0.02 & 505 & 1 & 0.98 & 0.18 \\
 &  & LP-P & 0.03 & 438 & \textbf{0} & 0.22 & 59.94 & 0.03 & 524 & \textbf{0} & 0.30 & 49.68 \\
\hline
\multirow{10}{*}{\shortstack[l]{Discontinuous \\ slope\\ change \\ known \\ variance}} & -- & Exact & 0.02 & 346 & 0 & 1.00 & 0.02 & 0.06 & 430 & 0 & 1.00 & 0.02 \\
\cline{2-13}
 & \multirow{3}{*}{50} & Bagel & 0.02 & 348 & \textbf{0} & \textbf{0.92} & \textbf{0.48} & 0.06 & 439 & \textbf{0} & \textbf{0.89} & \textbf{0.58} \\
 &  & LP-M & 0.02 & 351 & \textbf{0} & 0.87 & 1.62 & 0.07 & 442 & 2 & 0.79 & 2.99 \\
 &  & LP-P & 0.09 & 409 & \textbf{0} & 0.04 & 83.02 & 0.13 & 484 & \textbf{0} & 0.07 & 75.50 \\
\cline{2-13}
 & \multirow{3}{*}{100} & Bagel & 0.02 & 348 & \textbf{0} & \textbf{0.95} & \textbf{0.14} & 0.06 & 431 & \textbf{0} & \textbf{0.93} & \textbf{0.22} \\
 &  & LP-M & 0.02 & 348 & \textbf{0} & 0.91 & 0.69 & 0.06 & 435 & \textbf{0} & 0.86 & 1.35 \\
 &  & LP-P & 0.05 & 393 & \textbf{0} & 0.09 & 73.89 & 0.11 & 470 & \textbf{0} & 0.14 & 64.31 \\
\cline{2-13}
 & \multirow{3}{*}{200} & Bagel & 0.02 & 346 & \textbf{0} & \textbf{0.97} & \textbf{0.06} & 0.06 & 431 & \textbf{0} & \textbf{0.96} & \textbf{0.10} \\
 &  & LP-M & 0.02 & 348 & \textbf{0} & 0.94 & 0.30 & 0.06 & 431 & \textbf{0} & 0.91 & 0.58 \\
 &  & LP-P & 0.03 & 383 & \textbf{0} & 0.19 & 59.68 & 0.09 & 455 & \textbf{0} & 0.28 & 48.36 \\
\hline
\multirow{10}{*}{\shortstack[l]{Seasonal \\ change \\ known \\ variance}} & -- & Exact & 0.08 & 329 & 0 & 1.00 & 0.02 & 0.15 & 383 & 0 & 1.00 & 0.04 \\
\cline{2-13}
 & \multirow{3}{*}{50} & Bagel & 0.08 & 332 & 1 & \textbf{0.89} & \textbf{0.41} & 0.15 & 385 & \textbf{0} & \textbf{0.88} & \textbf{0.43} \\
 &  & LP-M & 0.08 & 335 & 3 & 0.80 & 1.47 & 0.15 & 390 & \textbf{0} & 0.77 & 2.29 \\
 &  & LP-P & 0.13 & 359 & \textbf{0} & 0.12 & 63.97 & 0.21 & 415 & \textbf{0} & 0.19 & 52.50 \\
\cline{2-13}
 & \multirow{3}{*}{100} & Bagel & 0.08 & 330 & \textbf{0} & \textbf{0.93} & \textbf{0.14} & 0.15 & 385 & \textbf{0} & \textbf{0.93} & \textbf{0.18} \\
 &  & LP-M & 0.08 & 329 & 1 & 0.87 & 0.66 & 0.15 & 386 & \textbf{0} & 0.85 & 0.96 \\
 &  & LP-P & 0.11 & 351 & \textbf{0} & 0.23 & 46.27 & 0.18 & 400 & \textbf{0} & 0.33 & 33.75 \\
\cline{2-13}
 & \multirow{3}{*}{200} & Bagel & 0.08 & 329 & \textbf{0} & \textbf{0.96} & \textbf{0.06} & 0.15 & 385 & \textbf{0} & \textbf{0.96} & \textbf{0.09} \\
 &  & LP-M & 0.08 & 330 & \textbf{0} & 0.91 & 0.32 & 0.15 & 385 & \textbf{0} & 0.92 & 0.39 \\
 &  & LP-P & 0.10 & 343 & \textbf{0} & 0.41 & 26.80 & 0.16 & 392 & \textbf{0} & 0.54 & 17.03 \\
\hline
\end{tabular}
}
\caption{Simulation result on a fixed horizon $T=2000$ with changepoint location $\tau=1000$ over 500 replications. Inv represents the invalid proportion, delay represents the average detection delay, MAP dist represents the median MAP distance, Jaccard represents the mean Jaccard index, and HPD err represents the HPD mass error. The lowest MAP distance, largest Jaccard index, lowest HPD error and lowest deviation between approximate coverage and exact coverage are in bold based on unrounded values. }
\label{appendix:example2}
\end{table}
\section{Computational results for Section 3.3}\label{Supplement A:sim_speed}
\tikzset{external/export next=false}
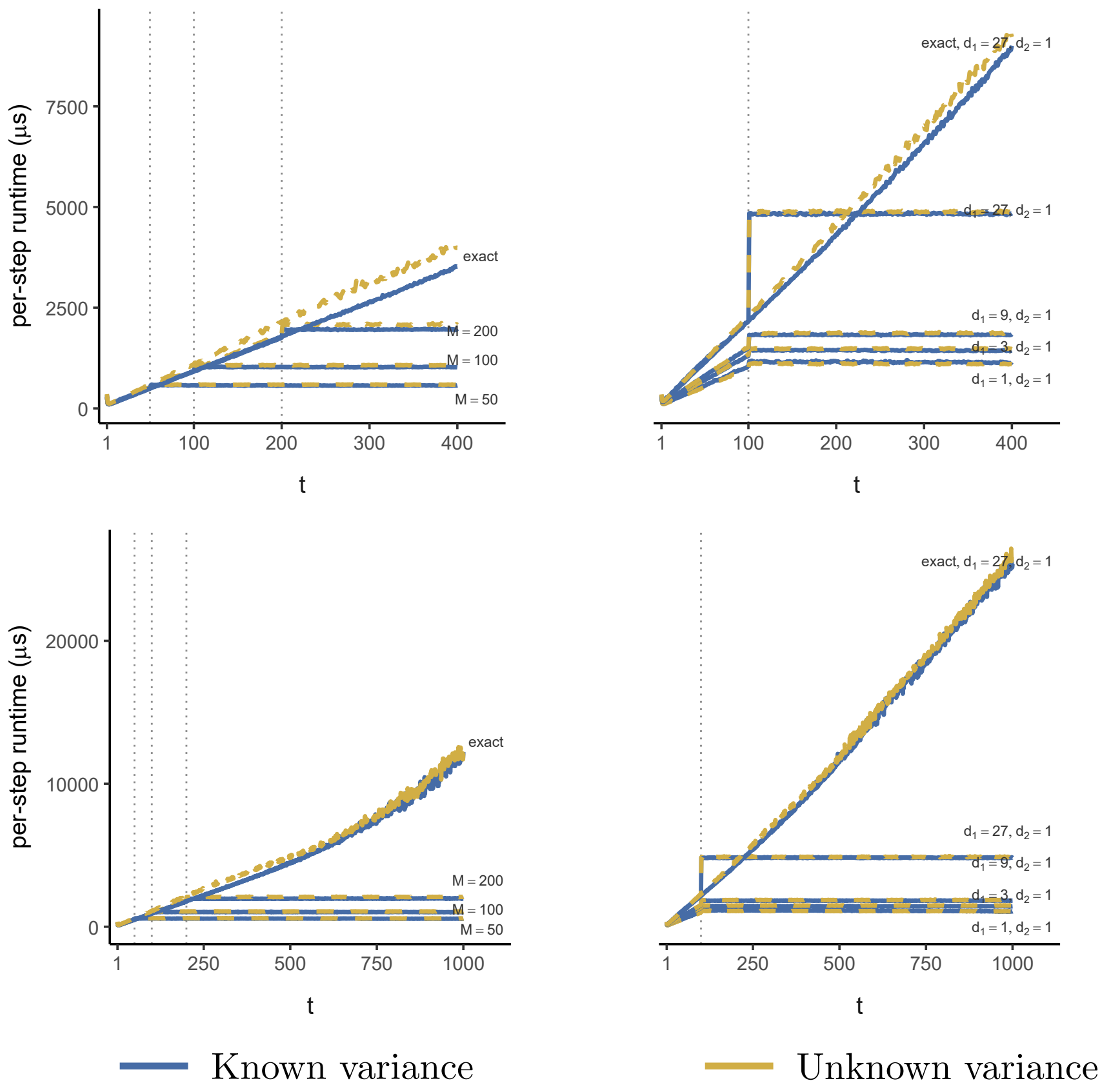
\begin{figure}[htp]
         \centering\includegraphics[width=\linewidth]{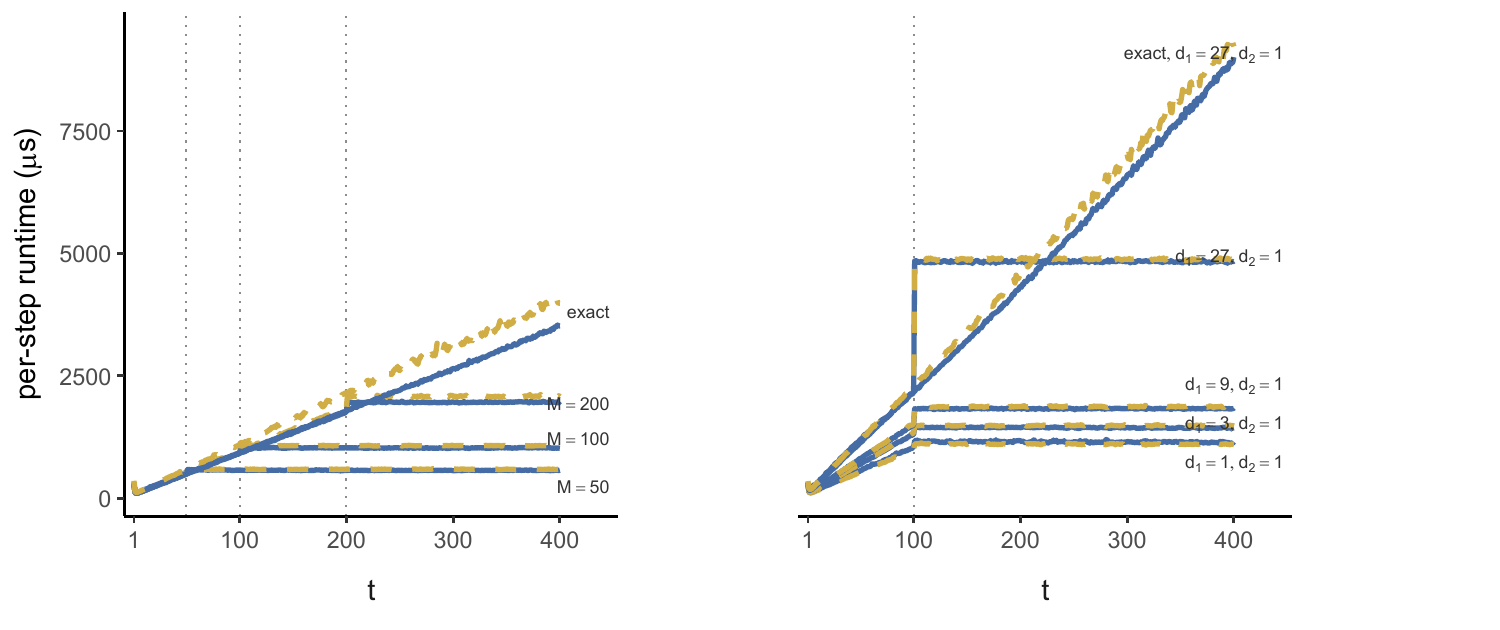}\\
         \centering\includegraphics[width=\linewidth]{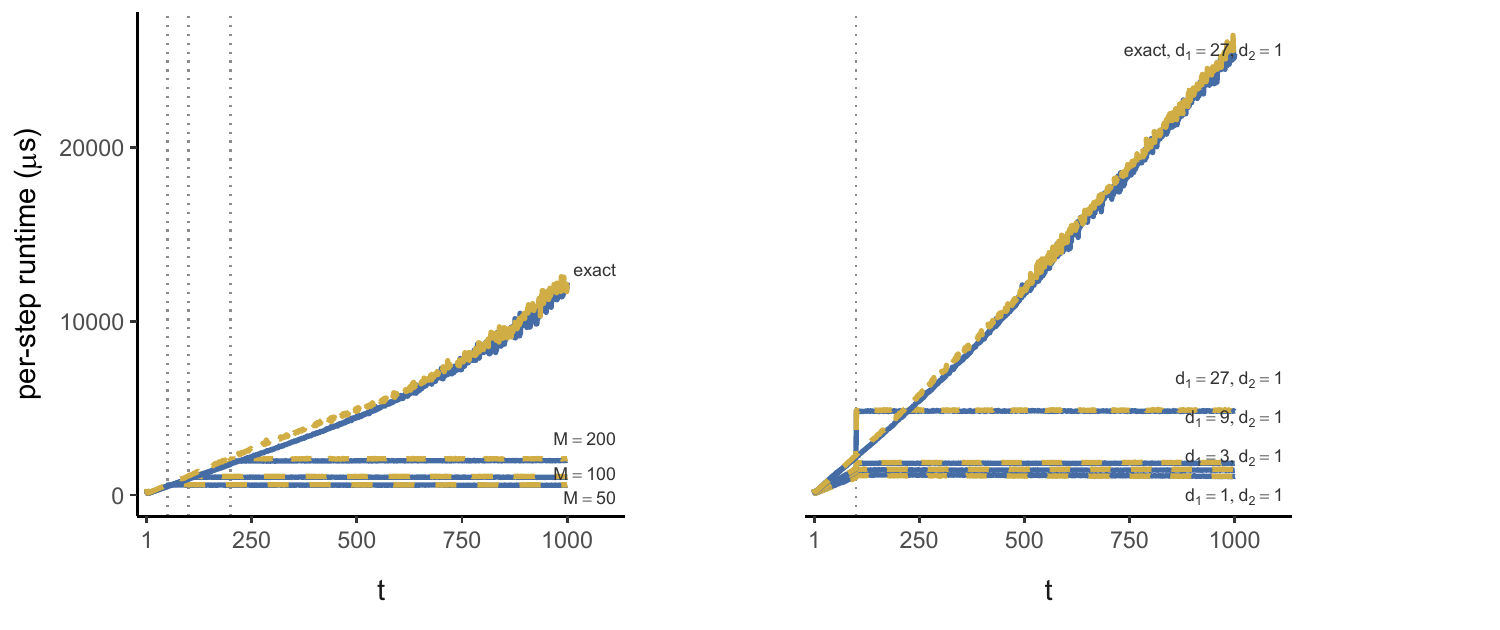}\\
        \begin{tikzpicture}
        \draw[
            line width=2pt,
            color={rgb,1:red,0.275;green,0.424;blue,0.651}
        ] (0.7,0.55) -- (1.4,0.55);
        \node[anchor=west] at (1.5,0.55) {Known variance};    
    \draw[
        line width=2pt,
        color={rgb,1:red,0.820;green,0.682;blue,0.271}
    ] (6.7,0.55) -- (7.4,0.55);
    \node[anchor=west] at (7.5,0.55) {Unknown variance};
        \end{tikzpicture}
    \caption{Median of per-step runtime of the \Bagel over 200 replications. The left figure shows the running time for different $M$ with fixed dimension of the design matrix $d_1=d_2=1$. The right figure shows the running time for different $(d_1,d_2)$ with fixed $M=100$. The fluctuation at the end of the exact approach is due to the memory footprint. The first row shows the result before $T=400$, and the bottom row shows the result before $T=1000$.}
    \label{fig:speed}
\end{figure}
Here we calculate the median of computational speed at each step of Bagel. The main code is implemented in C++, and it is invoked per step from R. The code was executed on a desktop machine with a 13th-generation Intel(R) Core(TM) i7, featuring 16 physical cores with a 3.4 GHz base clock, and 64 GB of RAM, running on Windows 11 and R 4.4.3 with a single thread. The running time is recorded through the system wall clock read immediately before and after each call to the recursion, so that the cost of every step is measured separately. 

Figure \ref{fig:speed} shows the median running time of \Bagel when varying the budget $M$ at a fixed dimension $d_1=d_2=1$ (left) and varying the dimension of the design matrix with fixed $M=100$ (right). The drop at the beginning is due to the warm-up period because the algorithm needs to initialise the R wrapper. The fluctuation at the end of the exact approach is caused by the memory footprint. Bagel can reduce an amount of computational cost of the exact approach, especially in the long run. The unknown variance costs are similar to the known variance case. 

There is a short period during which Bagel can be slower than the exact approach when the number of particles reaches its maximum length $M$. This is because Bagel needs to compute the distance and do additional matrix operations to determine which candidate to merge. After that, we can see that the per-step runtime of the exact approach continues to grow approximately linearly with $t$, but the per-step runtime of Bagel remains stable and constant. Therefore, the exact approach is more expensive than the approximated approaches over longer horizons.

\ifarxiv
\bibliographystyle{royal}
\bibliography{Ref}
\else
\bibliographystyle{ba2}
\bibliography{suppleARef}
\fi
\fi

\end{document}